\documentclass[%
 reprint,
 superscriptaddress,
nofootinbib,
 amsmath,amssymb,
 aps,
 prd,
 floatfix
]{revtex4-1}

\usepackage{graphicx}% Include figure files
\usepackage{dcolumn}% Align table columns on decimal point
\usepackage{bm}% bold math
\usepackage{amssymb}
\usepackage{color}
\usepackage[caption=false]{subfig}
\begin{document}

\preprint{APS/123-QED}

\title{BinarySkyHough: a new method to search for continuous gravitational waves from unknown neutron stars in binary systems}

\author{%
P.~B.~Covas$^{1}$, Alicia M. Sintes$^{1}$  %pep.covas
}\noaffiliation

\affiliation {Universitat de les Illes Balears, IAC3---IEEC, E-07122 Palma de Mallorca, Spain }% {BalearicIslands} {1}


\begin{abstract}
Non-axysymmetric, fast-spinning unknown neutron stars in binary systems may emit continuous gravitational waves (nearly-monochromatic long-duration signals) which can be detected by ground-based detectors like LIGO and Virgo. In this paper we present a new pipeline, called \textit{BinarySkyHough}, that can carry out all-sky searches for neutron stars in binary systems by exploiting the usage of graphics processing units. We give a detailed explanation of this new pipeline, and we present simulations which allow us to estimate the sensitivity of the new pipeline, which is approximately twice as sensitive as the best active pipeline with a comparable computational cost. %Up to date, only two pipelines which can perform this type of search exist, due to the massive computational cost that they posit.

%The sensitivity of the searches is limited by the available computational power, since the parameter space that needs to be searched is too big, and only one active pipeline currently exists. We present a new pipeline which is derived from an existing method to search for isolated neutron stars. We give a detailed explanation of this new pipeline, and we present several simulations which show that it has more than twice the sensitivity of existing methods at a comparable computational cost. We also present results for a veto and a follow-up method which have not been used before in searches for binary systems.

\end{abstract}

\maketitle

\section{\label{sec:intr}Introduction}

From the first and second observing runs of the Advanced LIGO and Virgo network, composed of three detectors called H1, L1, and V1, eleven detections of gravitational waves (GWs) from binary mergers have been reported \cite{Catalog}. Future observing runs may allow the detection of other types of sources, including gravitational waves emitted by non-axisymetric neutron stars (NS). Neutron stars are invaluable laboratories for physics under extreme conditions. These stars can emit continuous gravitational radiation  through a variety of mechanisms, including rotation with elastic deformations, magnetic deformations, unstable r-mode oscillations, and free precession, all of which operate differently in accreting and non-accreting stars (see \cite{Lasky} for a recent review). The detection of gravitational radiation from these sources would facilitate decisive progress on one of the most fundamental questions of modern science: the composition and state of matter at extreme densities. It would also allow one to test deviations from general relativity, for example polarization of the waves different from general relativity. % or deviation in the orbital evolution.

%Searches for CWs are usually split in three general categories, depending on the type of target and the amount of information that is known about it. 
Several searches for continuous gravitational-wave (CW) emission from fast-spinning galactic neutron stars, both isolated and in binary systems, have been developed and performed (see \cite{CWReview} for a recent review of CW searches). Depending on the type of target and the amount of information that is known, these searches are generally split in three categories: \textit{targeted searches} look for signals from known pulsars with known ephemerides; \textit{directed searches} look for signals from interesting sky positions (like the Galactic Center or Scorpius X-1), of which no information of the frequency evolution is known; \textit{all-sky searches} look for unknown neutron stars in our galaxy. Although none of these searches has detected CWs, interesting upper limits have been produced which already help to constrain some models of neutron star shape and equation of state \cite{O2KnownPulsars}.

Improvements in detector sensitivity, longer duration observing runs, and algorithmic improvements will all contribute to a compelling possibility for detection and observation of continuous gravitational waves, although there is still a large uncertainty on possible strength of continuous gravitational wave emission.

A very small percentage of the estimated neutron star population in our galaxy has been detected as pulsars \cite{NSPop}. We expect that these unseen neutron stars (perhaps with more extreme properties than the detected pulsar population) may emit detectable CWs, making all-sky searches a valuable endeavour. These searches need to calculate the Doppler modulation (produced by Earth's rotation and orbit around the Sun) for many sky positions,  which increases the complexity and computational cost of these searches. For this reason, the most sensitive methods (e.g. matched-filtering) cannot be used. Semi-coherent methods, which split the full observation time in smaller chunks and combine them by only tracking the frequency evolution of the signal (without taking into account the phase evolution between different chunks), are routinely used. To perform all-sky searches with a limited computational budget, semi-coherent methods have been proven to be more sensitive than coherent methods \cite{Cost}.

Approximately half of the known pulsars in the most sensitive frequency band of the ground-based detectors belong to binary systems. Neutron stars in binary systems have an additional modulation due to the NS movement around the binary barycenter (BB). Several directed searches for CWs from NS in known binary systems, such as Scorpius X-1, have been already performed \cite{O1DirectedBinary}. These searches usually have to deal with a four-dimensional parameter space comprised of the source frequency and three binary parameters, and also rely on semi-coherent methods to deal with the high computational cost.

%The extra modulation present in binary systems complicates even more the all-sky analysis by adding several more parameters which need to be taken into account, 
All-sky searches for NS in binary systems present an even harder problem, since the two sky positions need to be searched too, and the most sensitive semi-coherent methods used in all-sky searches for isolated systems cannot be used due to limited computational power. Currently, there are only two pipelines which can perform these searches, called \textit{TwoSpect} \cite{TwoSpectMethods} and \textit{NarrowBand Radiometer} \cite{AllSkyRadiometer}, and their sensitivity compared to the all-sky searches for isolated neutron stars is approximately 3 times worse. Only one all-sky search for NS in binary systems has been published until now \cite{TwoSpectResults} (by the \textit{TwoSpect} pipeline), having no detections and producing upper limits on the gravitational-wave amplitude. The lower sensitivity of these pipelines compared to all-sky searches for isolated NS calls for development of advanced techniques to improve the chances of detecting a CW signal from this type of system. %The lack of more pipelines and the lower sensitivity of the existing one calls for development of new techniques to improve the sensitivity of these searches. Up to now, only one search for CWs from unknown neutron stars in binary systems has been published \cite{TwoSpectResults}. 

In this paper we describe a new method to perform all-sky searches for CWs from NS in binary systems, which we call \textit{BinarySkyHough}. This method is an extension of the \textit{SkyHough} \cite{SkyHough} pipeline used for all-sky searches of isolated NS, and it benefits from the usage of GPUs (graphics processing units) in order to have a manageable computational cost. The plan of the paper is the following: section I gives background information on CWs and neutron stars; section II presents an overview of the type of signal that we search for; section III summarizes some important properties of the binary parameters of the pulsar population; section IV explains the new pipeline that we have developed; section V presents an estimation of the sensitivity of this new pipeline; section VI concludes with some final remarks.

\section{\label{sec:sigmodel}Signal model}

A neutron star with an asymmetry modelled as an aligned triaxial ellipsoid emits continuous gravitational waves, producing a time-dependent strain at the detectors given by \cite{Fstat}:
\begin{align}
    h(t) = h_0[&F_+(t,\psi, \hat{n})\frac{1+\cos{\iota}}{2} \cos{\phi(t)} \nonumber \\
    + &F_{\times}(t,\psi,\hat{n}) \cos{\iota} \sin{\phi(t)} ],
    \label{h0t}
\end{align}
where $F_+$ and $F_{\times}$ are the antenna patterns of the detectors (which can be found in \cite{Fstat}) for the two different gravitational-wave polarizations, $t$ is the time at the detector frame, the inclination angle $\iota$ is the angle between the neutron star angular momentum and the observer's sky plane, $\psi$ is the wave polarisation angle, $\phi(t)$ is the phase of the signal and $h_0$ is the amplitude of the signal given by:
\begin{align}
        h_0 = \frac{4\pi^2G}{c^4} \frac{I_{zz} \epsilon f^2}{d},
\end{align}
where $d$ is the distance from the detector to the source, $f$ is the gravitational-wave frequency (equal to two times the rotational frequency), $\epsilon$ is the ellipticity or asymmetry of the star, usually given by $(I_{xx}-I_{yy})/I_{zz}$, and $I_{zz}$ is the moment of inertia of the star with
respect to the principal axis aligned with the rotation axis. These two last quantities are related to the mass quadrupole $Q_{22}$ of the star:
\begin{align}
    \epsilon = \sqrt{\frac{8\pi}{15}} \frac{Q_{22}}{I_{zz}}.
\end{align}
%This model (aligned triaxial ellipsoid) assumes that the rotation axis is aligned with a principal axis of the star. If this is not true, another signal at once the rotational frequency is also emitted. For the rest of this paper, we will assume that the rotation axis is aligned with a principal axis of inertia. %We can see that the amplitude $h_0$ depends on the frequency which is not constant, but for the usual CW signals this change is very small and $h_0$ can be treated as constant.

%The phase of the gravitational-wave signal at the source frame is usually defined as a Taylor expansion in powers of frequency with respect to a referential time $\tau_{r}$ (where $\tau$ refers to times in the source frame). By definition, the gravitational-wave phase is locked to the rotational phase of the star, and from electromagnetic observations of pulsars we know that the rotational phase can be approximated by a Taylor expansion, since for most of them one or two frequency derivatives are enough to describe the detected electromagnetic pulses which are tied to the rotation of the star (following the pulsar lighthouse model). This phase is given by:
From electromagnetic observations of pulsars it is known that the rotational phase at the source frame can be described with a Taylor expansion in powers of frequency (usually one or two frequency derivatives are enough to describe the detected electromagnetic pulses, which are tied to the rotation of the star) with respect to a reference time $\tau_{r}$ (where $\tau$ refers to time in the source frame). Since the gravitational-wave phase is locked to the rotational phase, we can also describe the GW phase with a Taylor expansion:
\begin{align}
    \phi_s(\tau) = \phi_0 + 2\pi [f_0 (\tau-\tau_r) + \frac{f_1}{2!} (\tau-\tau_r)^2 + ...],
    \label{eq:phaseevosource}
\end{align}
where we define $f_0$ and $f_1$ as the frequency and first frequency derivative (spin-down/up), respectively, at $\tau_r$ and $\phi_0$ as an initial phase. To relate the phase in the source frame to the phase in the detector frame we need the timing relation between the source time and the detector time, which (neglecting relativistic wave propagation effects such as Einstein and Shapiro delays) is given by: % and assuming General Relativity is correct
\begin{align}
    %\tau + \frac{R(\tau')}{c} = t + \frac{\vec{r}(t')\cdot\hat{n}}{c} - \frac{d}{c},
    \tau + \frac{R(\tau)}{c} = t + \frac{\vec{r}(t)\cdot\hat{n}}{c} - \frac{d}{c},
    \label{eq:timing}
\end{align}
where $\vec{r}$ is the position of the detector with respect to the SSB, $\hat{n}$ is the position of the source in sky, and $R(\tau)$ is the radial distance of the binary orbit projected along the line of sight ($R>0$ means that the NS is further away than the BB, $R<0$ means otherwise). By neglecting the relativistic effects, we are assuming that the binary orbit can be described with a classical Keplerian orbit and that the relativistic effects such as the decay of orbital period do not produce a noticeable effect in our analysis. The decay of the orbital period is proportional to $P^{-5/3}$ (where $P$ is the orbital period), shorter periods producing faster decays. For a period of 0.01 days and a binary of stars with solar masses, the orbital decay is of the order of $10^{-11}$ s/s, which for observing runs of months or a few years would not affect our analysis.  %, but we remark that other pipelines which use longer coherent times such as also do not take into account these effects. %A brief discussion of this is presented in subsection \ref{subsec:RelEff}.
The projected radial distance of an ellipse is given by:
\begin{align}
    R &= a (1 - e \cos{E}) \sin{\iota_b} \sin{(\omega + \nu)} =  \nonumber \\
    &= a (1 - e \cos{E}) \sin{\iota_b} (\sin{\omega}\cos{\nu} + \cos{\omega}\sin{\nu}), %= \frac{a (1-e^2)}{1+e\cos{\Theta}},
    \label{eq:RadDist}
\end{align}
%where $\Theta=2\pi /P (\tau + \omega)$ is the true anomaly, $\omega$ is the angle of periastron passage, $P$ is the orbital period, $e$ is the eccentricity and $a$ is the semi-major axis amplitude.
where $\nu$ is the true anomaly, $e$ is the eccentricity of the orbit, $a$ is the semi-major axis amplitude, $\iota_b$ is the angle of inclination of the orbital angular momentum with respect to the observer's sky plane, the angle $\omega$ is the argument of periapsis and the angle $E$ is the eccentric anomaly, given by the transcendental equation:
\begin{align}
    \tau - \tau_p = \Omega (E - e \sin{E}),
    \label{eq:Etau}
\end{align}
where $\Omega=2\pi/P$ is the angular frequency, $\tau_p$ is the time of periastron passage (where $E=0$) and the true anomaly $\nu$ is related to the eccentric anomaly by:
\begin{align}
    \cos{\nu} = \frac{\cos{E} - e}{1 - e\cos{E}}.
    \label{eq:trueanomaly}
\end{align}

Combining equations \eqref{eq:RadDist} and \eqref{eq:trueanomaly} we have:
\begin{align}
    \frac{R(\tau)}{c} = a_p [\sin{\omega}(\cos{E} - e) + \cos{\omega} \sin{E}\sqrt{1 - e^2}],
    \label{eq:RC2}
\end{align}
where we have defined $a_p = a \sin{\iota_b} / c$. This equation depends on five orbital binary parameters: $a_p$, $\omega$, $\Omega$, $\tau_p$, and $e$, which fully describe the Keplerian elliptical orbit. 

%We want to search for binaries with very small eccentricity, so we will set $e=0$. In this case, the argument of periapsis $\omega$ becomes meaningless (it is not well defined) and we can set it to 0. The previous equation now reads:
If we choose to consider only orbits with small eccentricity ($e \ll 1$), the ELL1 model gives an approximation to lowest order in $e$ to equation \eqref{eq:RC2}, which is analytically tractable  \cite{ELL1}. This model substitutes the time of periastron passage $\tau_p$ with the time of passage through the ascending node $\tau_{\text{asc}}$, a quantity that remains well-defined even for circular orbits. These two times are related by:
\begin{align}
\tau_{\text{asc}} = \tau_p - \frac{\omega}{\Omega}.
\end{align}

For the ELL1 model the projected radial distance becomes:
\begin{align}
    \frac{R(\tau)}{c} = &a_p \sin[{\Omega(\tau-\tau_{\text{asc}})}] \nonumber \\
    + &a_p \frac{e\cos{\omega}}{2}\sin[{2\Omega(\tau-\tau_{\text{asc}})}] \nonumber \\
    + &a_p \frac{e\sin{\omega}}{2}\cos[{2\Omega(\tau-\tau_{\text{asc}})}] + \mathcal{O}(e^2).
    \label{eq:rc}
\end{align}

We want to express the phase evolution given by equation \eqref{eq:phaseevosource} in the detector frame. With equations \eqref{eq:timing} and \eqref{eq:rc} we can derive a relation between the times in the two different frames. The first step that we take is to redefine the constant reference time $\tau_r$ to $t_r$. We also drop the constant $d/c$ term, which is just changing again the reference time $t_r$ by a constant offset (this virtually joins the SSB and the BB). Now we can express $R(\tau)$ as $R(t_{SSB})$. We also redefine the constant $\tau_{\text{asc}}$ by $t_{\text{asc}}$, which again just changes the reference time by a constant factor. The timing equation now reads:
\begin{align}
    \tau = t + &\frac{\vec{r}(t)\cdot\hat{n}}{c} - a_p \sin[{\Omega(t-t_{\text{asc}})}] \nonumber \\
    & -a_p \frac{e\cos{\omega}}{2}\sin[{2\Omega(t-t_{\text{asc}})}] \nonumber \\
    & - a_p \frac{e\sin{\omega}}{2}\cos[{2\Omega(t-t_{\text{asc}})}].
\end{align}
With the previous equation, the phase model (without frequency derivatives in the source frame) at the detector frame now reads:
\begin{align}
    \phi(t) = \phi'_0 + 2\pi f_0 (t &- t_r + \frac{\vec{r}(t)\cdot\hat{n}}{c} - a_p \sin[{\Omega(t-t_{\text{asc}})}] \nonumber \\
    & - a_p \frac{e\cos{\omega}}{2}\sin[{2\Omega(t-t_{\text{asc}})}] \nonumber \\
    & - a_p \frac{e\sin{\omega}}{2}\cos[{2\Omega(t-t_{\text{asc}})}]).
    \label{eq:phaseMethod}
\end{align}

Since now we have the phase model in the detector frame, we can derive the frequency of the gravitational wave in the detector as:
\begin{align}
    f(t) = \frac{1}{2\pi}\frac{d\phi}{dt} = &f_0 + f_0\frac{\vec{v}(t)\cdot\hat{n}}{c} \nonumber \\
    - &f_0 a_p \Omega \cos{[\Omega(t-t_{\text{asc}})]} \nonumber \\
    - &f_0 a_p \Omega e \cos{\omega} \cos{[2\Omega(t-t_{\text{asc}})]} \nonumber \\
    + &f_0 a_p \Omega e \sin{\omega} \sin{[2\Omega(t-t_{\text{asc}})]}.
    \label{eq:frequencyevo}
\end{align}

Although the signal model for an eccentric orbit has been introduced, sensitivity estimations shown in section VI will assume orbits with zero eccentricity. If we assume that $e=0$, the frequency-time pattern is:
\begin{align}
    f(t) = f_0 + f_0\frac{\vec{v}(t)\cdot\hat{n}}{c} - f_0 a_p \Omega \cos{[\Omega(t-t_{\text{asc}})]},   \label{eq:frequencyevoFinal}
\end{align}
which depends on only three parameters: $\Omega$, $a_p$ and $t_{\text{asc}}$.

This model assumes that the neutron star does not suffer any glitches during the observing time, and that the effect of spin-wandering (stochastic variations on the rotational frequency due to the accretion process), if present, can be neglected.

\section{\label{sec:pulsars}Properties of the pulsar population}

\begin{figure}[tbp]
\includegraphics[width=1.0\columnwidth]{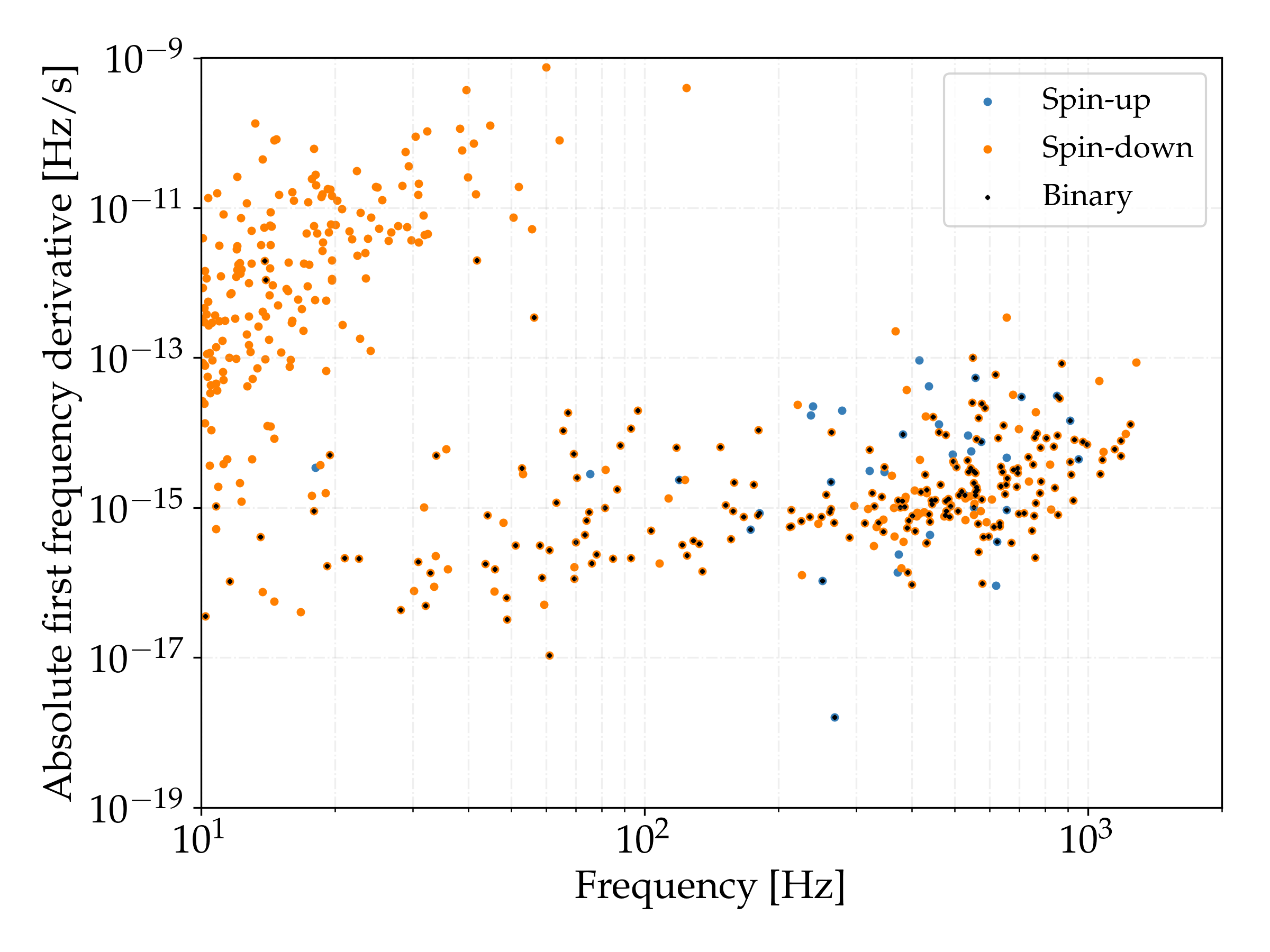}
\caption{\label{fig:p-pdot} Gravitational-wave frequency and absolute value of the gravitational-wave first frequency derivative for pulsars with gravitational-wave frequency greater than 10 Hz. Black dots indicate the pulsars which are part of a binary system. Data taken from \cite{ATNF} and downloaded with \cite{psrqpy}.}
\end{figure}

Estimations show that there might be around $10^8$ neutron stars in our galaxy, but only around $2700$ pulsars have been detected until now. From the known pulsars we can distinguish two different populations of neutron stars: the normal population and the millisecond population. The main difference between these two types is the rotational frequency and the spin-down/up, which can be seen in figure \ref{fig:p-pdot}: the millisecond pulsars spin much faster and have smaller frequency derivatives \cite{MSP}. These smaller frequency derivatives make the assumption of zero spin-down that was used to derive equation \eqref{eq:frequencyevoFinal} valid. % Better discussion on why this is: the resolution 1/Tc Tobs should be mentioned, and say that it is bigger that the spin-down values for millisecond pulsars

It is believed that most millisecond pulsars are recycled pulsars: they are or were part of a binary system, and they were spun up with accretion of matter from a companion. This process may explain the lower absolute value of the first frequency derivative of binary pulsars, since the accretion balances the rotational energy which is lost trough emission of electro-magnetic or gravitational waves \cite{MSP}. Furthermore, accretion of matter can bury the magnetic field of the NS. Decreasing the magnetic field lowers the electromagnetic energy emission, which makes the rotation more stable (since the rate of change of frequency is related to the energy loss), thus reducing their spin-down. This lowered electromagnetic energy can also explain why there have been more electromagnetic detections of normal pulsars than millisecond pulsars.

The accretion process can create and sustain quadrupole deformations which can be the source of CWs. Because the maximum observed rotational frequency is well below the maximum allowed by the limit imposed by the centrifugal break-up, it is believed that some process may be counteracting the neutron
star rotational acceleration before it reaches this maximum frequency. One proposed process is the emission of CWs. With this emission, neutron stars could reach a balance between accretion and emission of GWs, thereby sustaining a quadrupole which would emit CWs of amplitude given by the torque-balance limit, which can be estimated from the emitted x-ray flux for some NS like Scorpius X-1.

The majority of the detected pulsars which are supposed to emit CWs in the frequency band of the Advanced detectors (from 50 to 1000 Hz, approximately) are millisecond pulsars, as shown in figure \ref{fig:p-pdot}. Almost half of the millisecond pulsars belong to a binary system. For this reason, it is important to perform CW searches which take into account the different phase model which these signals have.

%Another difference between normal and millisecond pulsars is that while normal pulsars are usually in the disk of the galaxy (a region where lots of new stars are born), millisecond pulsars are more uniformly distributed around the halo of the galaxy, a zone where usually the oldest stars reside. This means that all-sky searches could be more useful for searches of NS in binary systems than isolated NS. %making all-sky searches more useful. 
%Millisecond pulsars are the oldest pulsars, 

%The eccentricity (defined by the division between the semi-major and semi-minor axis of the binary orbit) of pulsars in binary systems is shown in figure \ref{fig:eccfreq}. 

\begin{figure}[tbp]
%\begin{center}
\includegraphics[width=1.0\columnwidth]{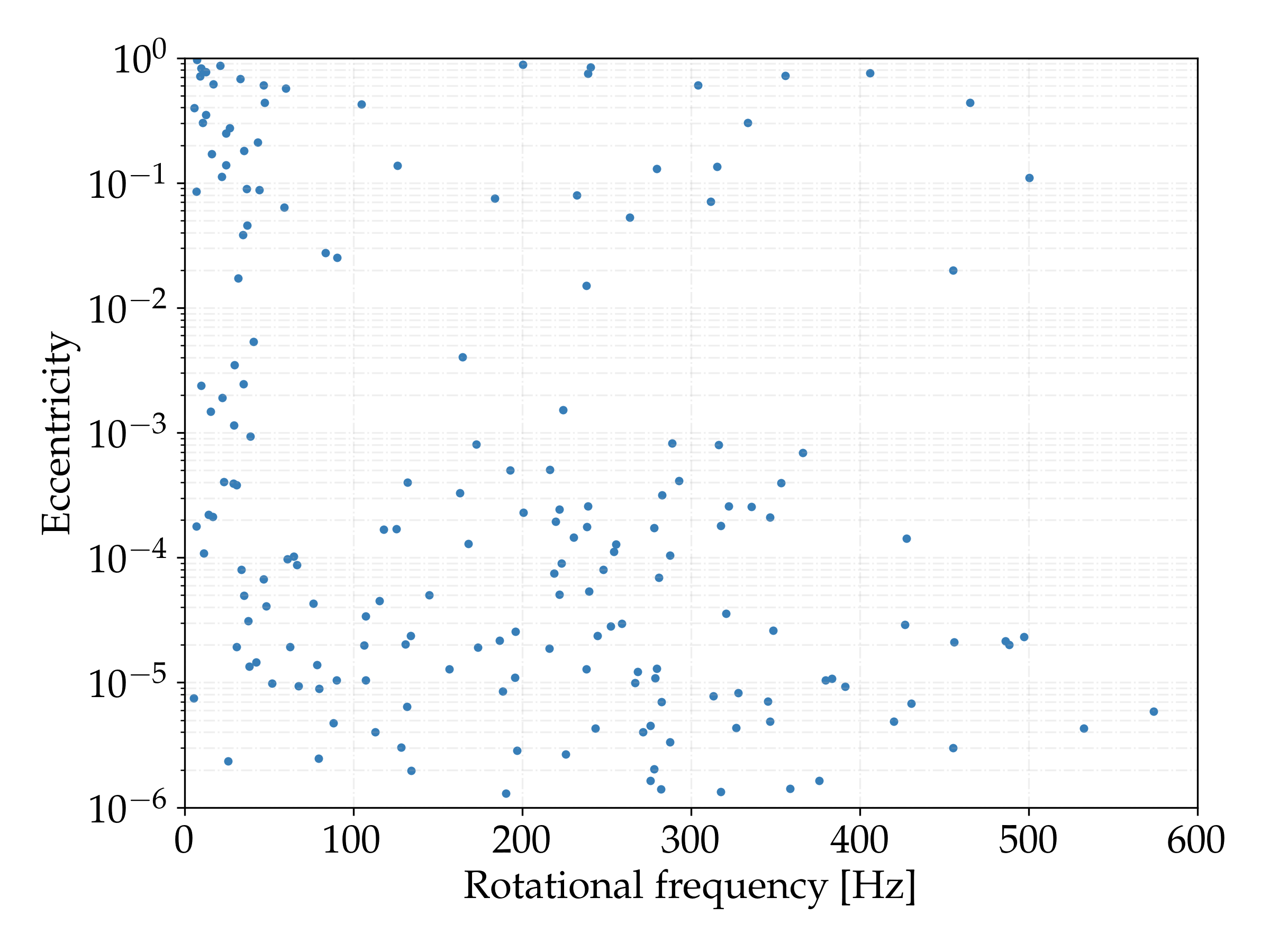}
\caption{\label{fig:eccfreq}Eccentricity of pulsars in binary systems as a function of their rotational frequency. Data taken from \cite{ATNF} and downloaded with \cite{psrqpy}.}
%\end{center}
\end{figure}

The eccentricity (defined as $\sqrt{1-b^2/a^2}$, where $a$ and $b$ are respectively the semi-major and semi-minor axis of the binary orbit) of pulsars in binary systems is shown in figure \ref{fig:eccfreq}. We observe that for most of the pulsars with measured eccentricity, it is smaller than 0.01 (for 167 out of 215). As we will see in section \ref{sec:sensitivity}, our pipeline is able to detect signals from systems with eccentricity up to 0.01 by using the zero-eccentricity model given by equation \eqref{eq:frequencyevoFinal}. %From this figure, we can see that our pipeline should be able to track %We also observe that there is no apparent correlation between the rotational frequency and the correspondent eccentricity. Plot with eccentricity vs $a_p$ and period.

%Figure \ref{fig:compmass} shows the measured median masses of the companion objects. The observed maximum amplitude of the Doppler shift is given by \cite{TwoSpectMethods}:
%\begin{align}
%    f_D = \frac{(2\pi G)^{1/3}}{c} \frac{f}{P^{1/3}} \frac{M_C}{(M_{NS}+M_C)^{2/3}} \sin{\iota_b},
%    \label{ObsDoppler}
%\end{align}
%where $M_{NS}$ is the mass of the NS and $M_C$ is the mass of the companion star. With our method we only can search for systems which produce an observable Doppler shift stronger than the width of a frequency bin. This fact limits the ability of our method to search for some astrophysical systems. Can we still load only 0.1 Hz bands? Yes.

%\begin{figure}[tbp]
%\includegraphics[width=1.0\columnwidth]{CompanionMass.png}
%\caption{\label{fig:compmass}.}
%\end{figure}

\begin{figure}[tbp]
%\begin{center}
\includegraphics[width=1.0\columnwidth]{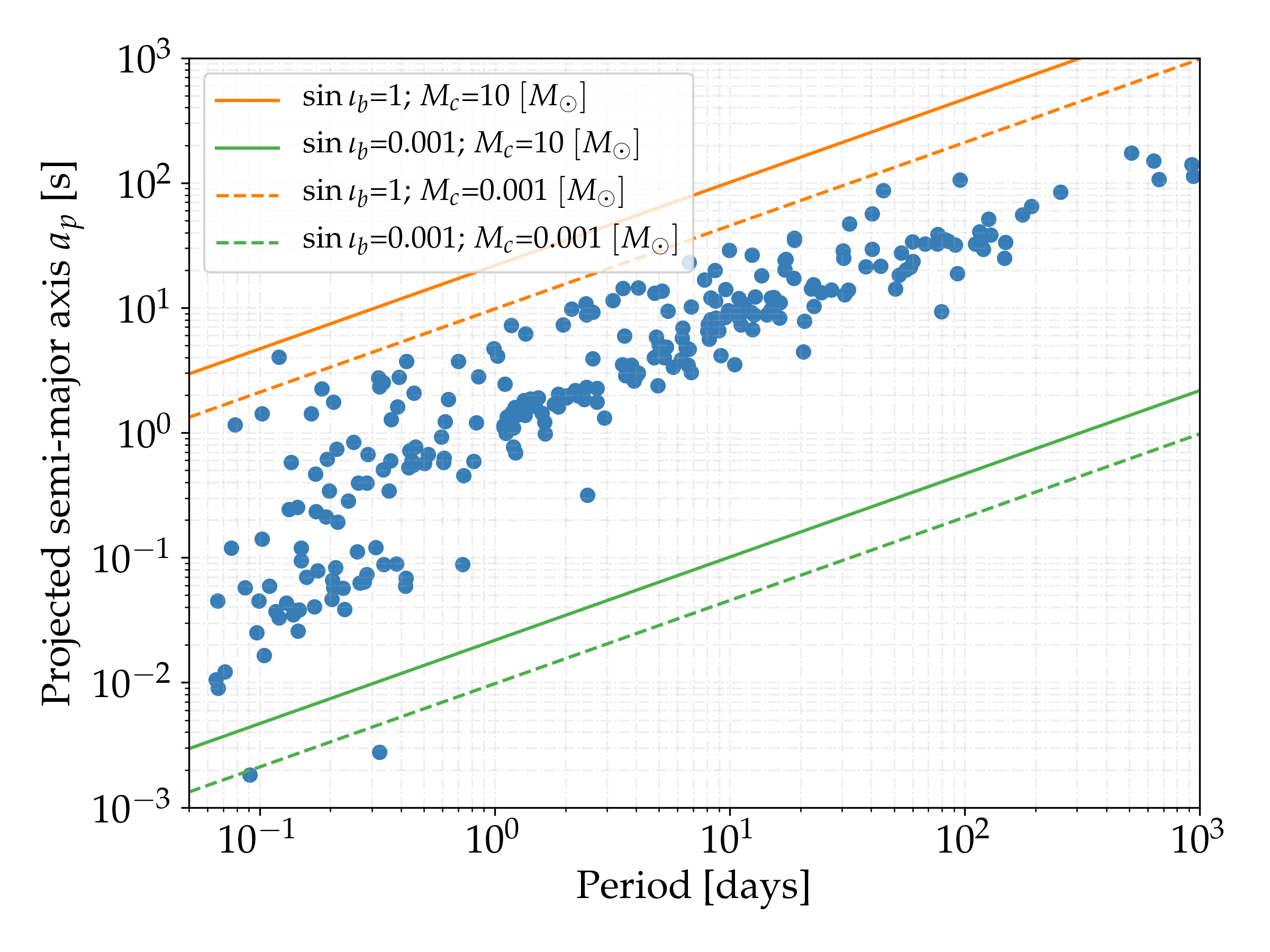}
\caption{\label{fig:P0-ap}The points show the projected semi-major axis amplitude and orbital period of known pulsars in binary systems. The four lines show different combinations of assumed companion mass and inclination angle for a $1.4$ $M_{\odot}$ neutron star, following equation \eqref{apP}. Data taken from \cite{ATNF} and downloaded with \cite{psrqpy}.}
%\end{center}
\end{figure}

For a Keplerian orbit, the projected semi-major axis amplitude $a_p$ and the orbital period $P$ follow the relationship given by the third Keplerian law:
\begin{align}
    %a_p = [\frac{G}{4\pi^2 c^3} (M_{NS} + M_C)]^{1/3} P^{2/3} \sin{\iota_b} - a_c,
    a_p = [\frac{G}{4\pi^2 c^3} (M_{NS} + M_C)]^{1/3} P^{2/3} \sin{\iota_b},
    \label{apP}
\end{align}
where $M_{NS}$ is the mass of the NS and $M_C$ is the mass of the companion star. We can see the values of these two quantities for the known pulsars in figure \ref{fig:P0-ap}. The different companion masses and angles of inclination $\iota_b$ account for the spread in the vertical axis. Along with the observational data, we have plotted four lines which follow equation \eqref{apP} for two different values of the companion mass and two values of the inclination angle (for a $1.4$ $M_{\odot}$ neutron star). These observational points from known pulsars can guide the choice of parameter space that we want to search, as we will discuss later. %For example selecting a certain range of orbital period gives a limited range for the choice of $a_p$ and completely constrains $t_{asc}$. %We could restrict a search to follow the main trend of this figure, but as we can see by the lines maybe there is an observational bias and we don't detect certain portions of parameter space.

\section{\label{sec:hough}SkyHough method for isolated searches}

The \textit{SkyHough} semi-coherent method is detailed in \cite{SkyHough}. This method based on the Hough transform is used to perform all-sky searches of CWs from isolated neutron stars. It exploits the Doppler modulations produced by Earth's movement around the Solar System Barycenter (SSB) by reusing the same Doppler modulation for several frequency bins in order to save computational costs, which makes the \textit{SkyHough} pipeline the cheapest semi-coherent method to track modeled signals currently available and the best choice to be adapted for a  search for NS in binary systems.

\begin{figure*}[tbp]
%\begin{center}
\includegraphics[width=1.0\linewidth]{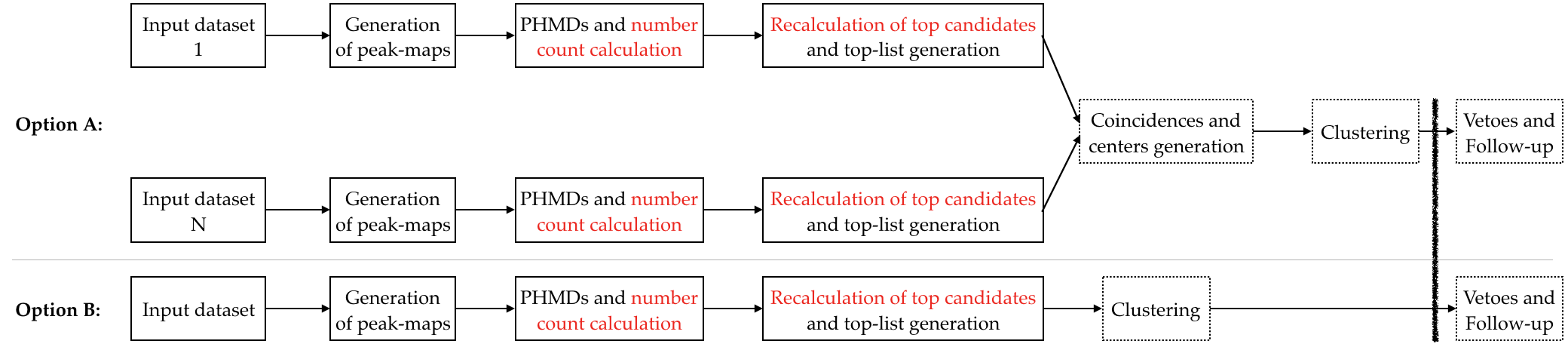}
\caption{Flowchart showing the steps of the \textit{BinarySkyHough} algorithm for one search band. Red text shows the steps performed with GPU kernels, and dotted boxes represent the post-processing stages (different vetoes and follow-up procedures are not part of the \textit{BinarySkyHough} pipeline). Option A shows the steps followed when the data is divided in $N$ different datasets, and option B shows the steps when the data is not divided. The main difference is that option B does not apply the coincidences step in the post-processing stage.}
\label{fig:flowchart}
%\end{center}
\end{figure*}

Figure \ref{fig:flowchart} shows a flowchart which summarizes the different steps of the method. We will discuss in detail these steps in the next subsections. As can be seen in this figure, this pipeline can be run with two different options: option A divides the data in $N$ different datasets, while option B does not. The main difference is that option B does not apply the coincidences step in the post-processing stage as will be detailed in section V.E.

\subsection{Input data}
\textit{SkyHough} starts by splitting the data from an observation run into smaller chunks of duration $T_c$ and then produces Short Fourier Transforms (SFTs) from the calibrated and windowed $x(t)$ data produced by the different ground-based detectors such as H1, L1 or V1. The choice of length of these SFTs is detailed in \ref{subsec:CohTime}. 

The data $x(t)$ obtained by the detectors is given by:
\begin{align}
    x(t) = h(t) + n(t),
\end{align}
where $h(t)$ is given by equation \eqref{h0t} and $n(t)$ is the noise present in the detector, which is related to the one-sided power spectral density (PSD) by:
\begin{align}
    S_n (f) = 2 \int_{-\infty}^{\infty} \langle n(t)n(0) \rangle e^{-2\pi i f t}dt,
\end{align}
where $\langle . \rangle$ denotes an ensemble average. The discrete Fourier Transforms are defined by:
\begin{align}
    \tilde{x}_k = \Delta t \sum_{j=0}^{M-1} x_j e^{-2\pi i j k / M},
\end{align}
where $j$ is a timestamp index, $k$ a frequency bin index, $M=T_c/\Delta t$ where $\Delta t$ is the inverse of the sampling frequency.

The dataset to be analyzed is split in a number $N_{SFTs}$ of different chunks, which are used to produce a spectrogram (a matrix of frequency-time bin powers). These powers are afterwards normalized:
\begin{align}
    \rho_{Jk} = \frac{|\tilde{x}_{Jk}^2|}{\langle n_{Jk} \rangle^2},
    \label{eq:HoughPower}
\end{align}
where $J$ is the SFT index and $\langle n_{Jk} \rangle^2$ is usually estimated with a running median over a number of frequency bins (usually 101 bins are used). The estimated noise is related to the PSD by:
\begin{align}
    \langle n_{Jk} \rangle^2 \approx \frac{T_c}{2} S_{n;J} (f_k).
\end{align}

% AR estimation, use of cleaned data, local maxima for peak selection.

%The optimal choice for the power threshold $\rho_{th} = 1.6$ was found in \cite{SkyHough}, although it was obtained for an unweighted implementation of the Hough pipeline.

The spectrogram is replaced by 1s and 0s by defining a power threshold $\rho_{th}$ (if the power in a bin is above this threshold, it is substituted by a 1, otherwise it is substituted by a 0). In \cite{SkyHough} it was found that for a signal embedded in Gaussian noise (without taking into account the weights) the optimal choice for this threshold is $\rho_{th} = 1.6$. These 1s and 0s are multiplied by per-SFT weights (calculated at the mid-time of each SFT), given by:
\begin{align}
    w_J \propto \frac{F_{+;J}^2 + F_{\times;J}^2}{S_{n;J}},
    \label{eq:weigths}
\end{align}
which give more importance to times when the data has lower noise and when the detectors are optimally oriented to the specific sky position being searched. These weights were derived in \cite{SkyHoughWeights}.

%After calculating the weight for each SFT, all the weights are multiplied by a normalisation factor:
%\begin{align}
%    w_i = \frac{N_{SFTs}}{\sum_{i=1}^{N_{SFTs}}w'_i} w'_i.
%    \label{eq:weightnorm}
%\end{align}

\subsection{Partial Hough map derivatives and look-up table approach}
The \textit{SkyHough} pipeline calculates the so-called ``Partial Hough Map Derivatives" (PHMDs) at each timestamp and frequency bin \cite{SkyHough}. These structures contain the weighted 1s and 0s and are calculated by using the fact that at a given time, a circle of sky positions produces the same Doppler modulation, given by:
\begin{align}
    \cos{\theta} = \frac{ \vec{v}(t) \cdot \hat{n} } { v(t) } = \frac{c}{v} \frac{f(t) - \hat{f}(t)}{\hat{f}(t)},
    \label{eq:hough1}
\end{align}
where $\theta$ is the angle between $\vec{v}$ (the velocity vector of a detector) and $\hat{n}$ (the position of the source on the sky). 

Due to the limited resolution in frequency, the group of sky positions producing the same modulation is given by an annulus (centered on the velocity vector of the given detector) of a certain width $\Delta \theta$ instead of a circle:
\begin{align}
\cos{\Delta \theta} = \frac{c}{v} \frac{n \delta f}{\hat{f}(t)} = \frac{n }{n_0},
\end{align}
where $\delta f$ is the width of a frequency bin and $n$ is a number which indexes the frequency modulation, from 0 to a maximum of $n_0=f v/(c \delta f)$.  At a given time and for an observed frequency, all the sky is covered by a finite amount $n_0$ of annuli produced by the Doppler modulation. %The maximum number of frequency bins is:

Each PHMD contains all the possible annuli for a certain sky-patch (set of sky positions) compatible with a given time and observed frequency. In other words, one sky position will produce different modulations at different timestamps, and for this reason it will be mapped to different  frequency bins at different timestamps.

\begin{figure}[tbp]
\includegraphics[width=1.0\columnwidth]{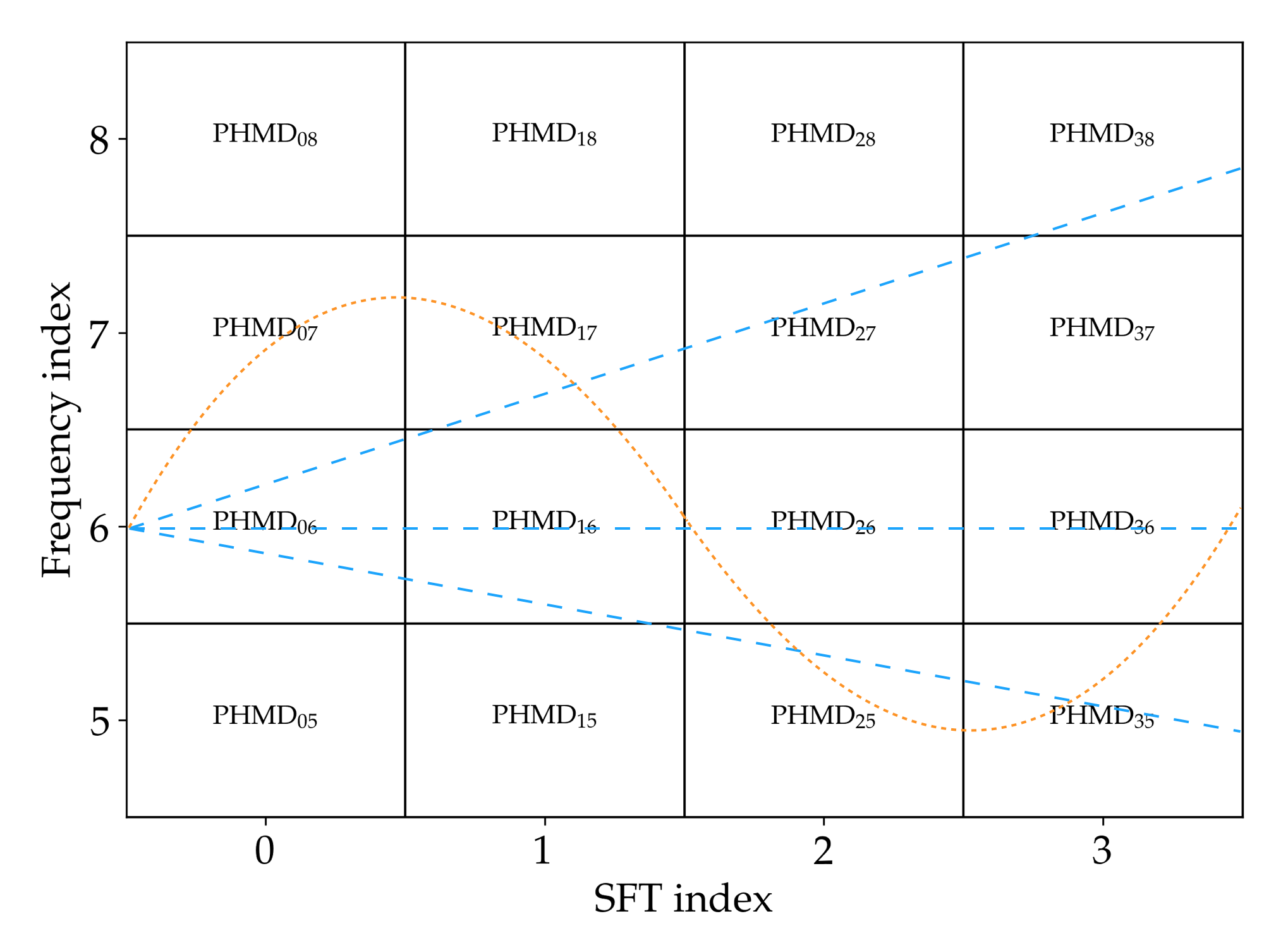}
\caption{\label{fig:PHMD}An example showing the PHMD scheme for the Hough pipelines. The dotted line shows the type of tracks which the \textit{BinarySkyHough} pipeline uses to combine PHMDs at different timestamps, while the dashed lines show some tracks given by different spin-down/up parameters, which are used by the \textit{SkyHough} pipeline.}
\end{figure}

After calculating all the PHMDs, as pictured in figure \ref{fig:PHMD}, the pipeline adds one PHMD for each timestamp by following the frequency path created by the source frequency variation (given by $f_0 + f_1t$), which is the change produced by the spin-down/up of the source. This produces a final Hough map for each combination of $f_0$, $f_1$ and sky-patch. %This is shown schematically in figure \ref{fig:PHMD}.

The usage of the PHMDs greatly reduces the computational cost of the pipeline since many sky positions are analyzed at once (the ones in the sky-patch), because only the borders of the selected annulus need to be tracked. %and only those whose annulus is coincident with a frequency bin which has been substituted by a 1 are taken into account in the loop which calculates the final Hough map.

Furthermore, \textit{SkyHough} uses the look-up table (LUT) approach. This method reuses the same Doppler modulation for contiguous searched frequency bins. This means that it calculates the PHMDs once instead of calculating them for each searched frequency. The same frequency which appears in the denominator of equation \eqref{eq:hough1} is used for a number of  searched frequencies (up to a maximum, given by equation 4.24 of \cite{SkyHough}), changing only the starting frequency $f_0$ to calculate the source frequency variation. The LUT approach produces computational savings in exchange of searching for modulations which are not exact, which lowers the sensitivity of the search (a quantitative estimation of the effect of the LUT in sensitivity doesn't exist).

\subsection{Detection statistics}
The analysis of the full parameter space is split in smaller frequency bands and in sky-patches of size dependent upon frequency. For each of these regions a top-list is produced, which contains information about the most significant candidates in that region. The analysis is divided in two steps with different detection statistics, as explained below.

The detection statistic which ranks the searched templates in the first step is the number count significance, given by:
\begin{align}
    s_H = \frac{n - \langle n \rangle}{\sigma_H},
    \label{eq:hough2}
\end{align}
where $\langle n \rangle$ and $\sigma_n$ are the expected mean and standard deviation of the Hough number count $n$ (the weighted sum of 1s and 0s summed along the frequency-time pattern of the signal) when only noise is present, given in \cite{SkyHough}. %This detection statistic is used to rank all the searched templates and put them ordered by significance in a top-list.

After calculating the number count significance for all the templates using the LUT approach in the first step, a selected number of best templates (a given fraction $N_C$ of the total, e.g. $N_C=0.1$ or 10\%) are passed to a second step of the search for which a different detection statistic is calculated by using the exact frequency path. This detection statistic is the power significance, given by:
\begin{align}
    s_{\mathcal{P}} = \frac{\mathcal{P} - \langle \mathcal{P} \rangle}{\sigma_{\mathcal{P}}},
    \label{eq:hough3}
\end{align}
where $\mathcal{P}$ is the sum of weighted normalized powers given by equation \eqref{eq:HoughPower}. This detection statistic is more sensitive than the number count significance since it encodes the full power information contained in the original data, without thresholding to 1s and 0s. 

%If $N_C$ is small enough (of the order of 0.01 or less), the computational cost of performing the two steps of the search should be smaller than performing a search with a unique step which calculates the $s_P$ detection statistic, but with an almost equal sensitivity. 

The second step of the search is performed on the $N_C$ best templates returned by the first step. This second step reorders these $N_C$ templates, which are then cut to a smaller number (i.e. $N_{CF} \ll N_{C}$) since returning $N_C$ candidates for each searched region would be impossible in terms of the size of output files. The final top-list returned by the pipeline is the set of $N_{CF}$ templates ordered by $s_{\mathcal{P}}$. %The number of templates in this top-list $N_{FC}$ is smaller than $N_C$. This means that the second step of the search performs a reordering of the templates from the first step, which are then cut to a smaller number (i.e. $N_{CF} \ll N_{C}$).

%Of course, the lower $N_C$ is, the smaller the contribution of the second step will be.

In this way, the first step of the search serves to find interesting regions in parameter space, which are then analyzed with greater sensitivity at the second step. This procedure allows to recover a large fraction of the lost sensitivity due to the usage of the LUT approach and a less sensitive detection statistic in the first step, with a computational cost much smaller than running the second step directly for all the initial templates (if the number of templates passed to the second step is more than an order of magnitude smaller than the total number of templates).

Furthermore, at the second step of the search more SFTs are used than in the first step. A complementary set of SFTs is generated from the initial set, by moving the initial time of each SFT by $T_c/2$ and creating new SFTs at each new timestamp (if a contiguous set of data of $T_c$ seconds exists). This procedure improves the sensitivity of the second step, and follows the ideas outlined in \cite{Sliding}.

%This is achieved by sliding the initial times of each SFT that was used at the first step, obtaining more SFTs (approximately twice the previous amount), all of them of $T_c$ contiguous seconds.

%The final top-list returned by the pipeline are the templates ordered by the power significance. The number of candidates which are passed from the first step of the search to the second one is a tuneable parameter called $N_{C}$, which should be a high enough percentage of the total number of templates. This second statistic allows to recover a fraction of the lost sensitivity in the first step due to the usage of the LUT approach.

%Furthermore, we apply windowing to the data to reduce leakage. SFTs generation. Double SFTs at second step.

%Peakmap generation, AR estimation, peak threshold optimal choice.

\section{\label{sec:BSH}New method for CW all-sky binary searches}

In this section we present \textit{BinarySkyHough}, a new pipeline to perform all-sky searches of CWs from unknown neutron stars in binary systems with low-eccentricity orbits. The new method is an adaptation of the \textit{SkyHough} semi-coherent method \cite{SkyHough} which has been used for all-sky searches of CWs from isolated neutron stars like \cite{S2Hough,S4CWAllSky,S5Hough,O1CWAllSkyFull}. \textit{SkyHough} is the semi-coherent pipeline with lowest computational cost, with a sensitivity similar to other methods that are an approximately an order of magnitude more computationally costly \cite{MDC,O2AllSky}. This makes it an excellent candidate to be adapted to perform CW searches in binary systems, which have a computational cost higher than isolated searches due to the extra parameters that need to be searched.

\subsection{BinarySkyHough}

The \textit{SkyHough} pipeline calculates the PHMDs and then sums them following the path given by $f_0 + f_1 t$. To implement a search for neutron stars in binary systems, we substitute the search over spin-down/up values with a search over binary parameters. This change is shown in figure \ref{fig:PHMD}. As we saw in section \ref{sec:pulsars}, the frequency derivative of pulsars in binary systems usually is smaller than isolated NS and this parameter doesn't need to be searched for. This means that the emission of energy from the neutron star is balanced by accretion, which would produce a bright X-ray emission (as observed for Sco-X1), although this might not be observable due to misalignment from the emission spot to the line of sight or due to screening from other objects.

At this point in development we only allow circular orbits (zero-eccentricity), which are described by three parameters: $\Omega=2\pi/P$, $a_p$ and $t_{\text{asc}}$. As equation \eqref{eq:frequencyevoFinal} shows, this model consists of six different parameters: the initial frequency $f_0$, the sky position given by the right ascension $\alpha$ and declination $\delta$, and the three binary parameters. Comparing with the isolated case, we go from a 4-dimensional parameter space to a 6-dimensional one. Although we assume the eccentricity of the system is 0, this pipeline remains sensitive to systems with eccentricity $e < 0.01$ without the need to explicitly search over it, as will be explained in sections \ref{subsec:MaxEcc} and \ref{sec:sensitivity}.%All the other steps of the method are applied without further changes.

The innermost loop over the binary parameters, which calculates the frequency-time path and sums the different PHMDs, has been ported to CUDA (a parallel computing platform which allows to control GPUs) in order to take advantage of the massive parallelism that GPU cards provide. Section V.F shows a comparison of timings for different runs, and we can see that without the GPUs this search would take an unfeasible time to run. The final loop for each sky-patch, which calculates the second detection statistic for a limited number of templates, has also been ported to CUDA to further speed-up the code.

\subsection{\label{sec:resolution}Resolution, parameter space range and number of templates}

\subsubsection{Resolution}
In order to construct the template bank which contains the templates that are going to be searched, we need to define the resolution of parameter space which decides the spacing between templates. The usual metric which quantifies the needed resolution is the mismatch, which gives one minus the ratio of recovered signal-to-noise ratio SNR$_r$ to the SNR which would be obtained if the matching was performed with a template using the true signal parameters (when noise is not present in the data). The mismatch is given by:
\begin{align}
    \mu_0 = \frac{\text{SNR}^2 - \text{SNR}_r^2}{\text{SNR}^2},
    \label{eq:mismatch}
\end{align}
which is a number between 0 (fully recovered SNR) and 1 (no recovered SNR). To estimate this mismatch, an approximation which is usually used is the phase metric \cite{PrixMultiMetric}:
\begin{align}
    \mu_0 \approx g_{ab} (\lambda) d\lambda^a d\lambda^b,
    \label{eq:phasemetric}
\end{align}
where $g_{ab}$ is the parameter space metric ($a$ and $b$ run over the dimensions, given by the number of parameters) and $\lambda$ represents the different parameters (e.g. frequency, sky position, ...). This approximated mismatch is unbounded and can be higher than 1, and from previous studies it is known that this approximation highly overestimates the actual mismatch for mismatches higher than $0.5$ \cite{PrixMultiMetric}. %is valid for $T_{obs}>1$ days
The phase metric is calculated as:
\begin{align}
    g_{ab} = \langle \delta_a \phi (\lambda) \delta_b \phi (\lambda) \rangle - \langle \delta_a \phi (\lambda) \rangle \langle \delta_b \phi (\lambda) \rangle,
    \label{eq:phasemetric2}
\end{align}
where the factors inside $\langle . \rangle$ are averaged over different SFTs during the observing time, and $\phi$ is given by equation \eqref{eq:phaseMethod}.

%One of the main questions that needs to be answered is: what grid spacing do we need to in order to recover  the astrophysical signals? The theoretical procedure uses the metric, obtained by allowing a maximum mismatch between searched templates and real signals.

%It is known that for mismatches bigger than 0.5, the predicted mismatches are overestimating the real mismatch (other assumptions, maximum obs time, combination of hough and phase metric resolutions, ...).

%Frequency variation equations

The metric for directed binary searches was obtained in \cite{PaolaMetric} and in \cite{WhelanMetric}. %, where it was tested and estimations of its correctness were shown. 
We will use the equations obtained in \cite{PaolaMetric} for the semi-coherent short-segment regime (equations 62), where $T_c \ll P$, which sets a lower limit for the orbital periods that we will be able to search. The resolution for the binary parameters is:
\begin{align}
    \delta a_p &= \frac{\sqrt{6m_a}}{\pi T_c f \Omega} \\
    \delta \Omega &= \frac{\sqrt{72 m_{\Omega}}}{\pi T_c f a_p \Omega T_{obs}} \\
    \delta t_{\text{asc}} &= \frac{\sqrt{6 m_t}}{\pi T_c f a_p \Omega^2},
    \label{eq:binaryres}
\end{align}
where $m_x$ are the mismatch parameters, which quantify the amount of lost SNR and define the desired spacing between templates. These equations were obtained for a coherent detection statistic, which instead of tracking the frequency-time evolution tracks the full $h(t)$ evolution given by equation \eqref{h0t}. For this reason, in our case these mismatch parameters $m_x$ will not correspond to an actual mismatch value and will only represent a tuneable parameter in our pipeline. %As we will use these equations for a semi-coherent method which does not try to match the phase, we remark that the mismatch parameters $m_x$ do not correspond to an actual mismatch value and that they will only represent a tuneable parameter in our pipeline. %We have checked that the scalings given by these equations are correct.

It can be seen that these equations depend on the region of binary parameter space for which they are calculated, as opposed to the resolution for the spin-down/up which only depends on the coherent time and the observing time \cite{SkyHough}. For shorter periods (greater $\Omega$) the resolution is increased for all three parameters, and the parameter space separation between templates is reduced. The total number of templates will be proportional to $\Omega^4$, which highly complicates the feasibility of the analysis for short periods.

In \cite{PaolaMetric}, the resolution was obtained for a directed search, which does not search over sky positions as all-sky searches do. For the frequency and sky positions we will use the resolutions obtained for the isolated \textit{SkyHough} pipeline \cite{SkyHough}:
\begin{align}
    \delta f &= \frac{1}{T_c} \\
    \delta \Theta &= \frac{c}{v T_c f P_F},
\end{align}
where $\Theta$ represents any of the two sky positions and the pixel-factor $P_F$ is another tuneable mismatch parameter.

%The frequency resolution can be compared to the frequency resolution of \cite{PaolaMetric}:
%\begin{align}
%    d'f &= \frac{\sqrt{3m}}{\pi T_c},
%\end{align}
%which has the same scaling with the coherent time and with a particular choice of the mismatch parameter $m\approx 3.3$ is exactly the same.

%The sky resolution can be compared with other resolutions obtained by the phase metric method like X:
%\begin{align}
%    d'\theta &= \frac{\sqrt{m}}{0.021 \pi f},
%\end{align}

We have verified the scalings given by these equations through extensive simulations (shown in section \ref{sec:sensitivity}), different mismatch values and across different regions of $\Omega$ and $a_p$. %A more in-depth study of the binary metric and mismatch for all-sky semi-coherent searches is left for future work. %Since we are mixing resolutions obtained with different methods, we don't claim that the mismatch parameters appearing in equation \eqref{eq:binaryres} truly correspond to the SNR lost when trying to match the signal. We simply treat them as a scaling parameter, and we study with simulations the behaviour of our pipeline for different numbers and combinations. A more in-depth study of the binary metric and mismatch for all-sky semi-coherent searches is left for future work.

\subsubsection{Range of parameter space}
\begin{figure}[tbp]
\includegraphics[width=1.0\columnwidth]{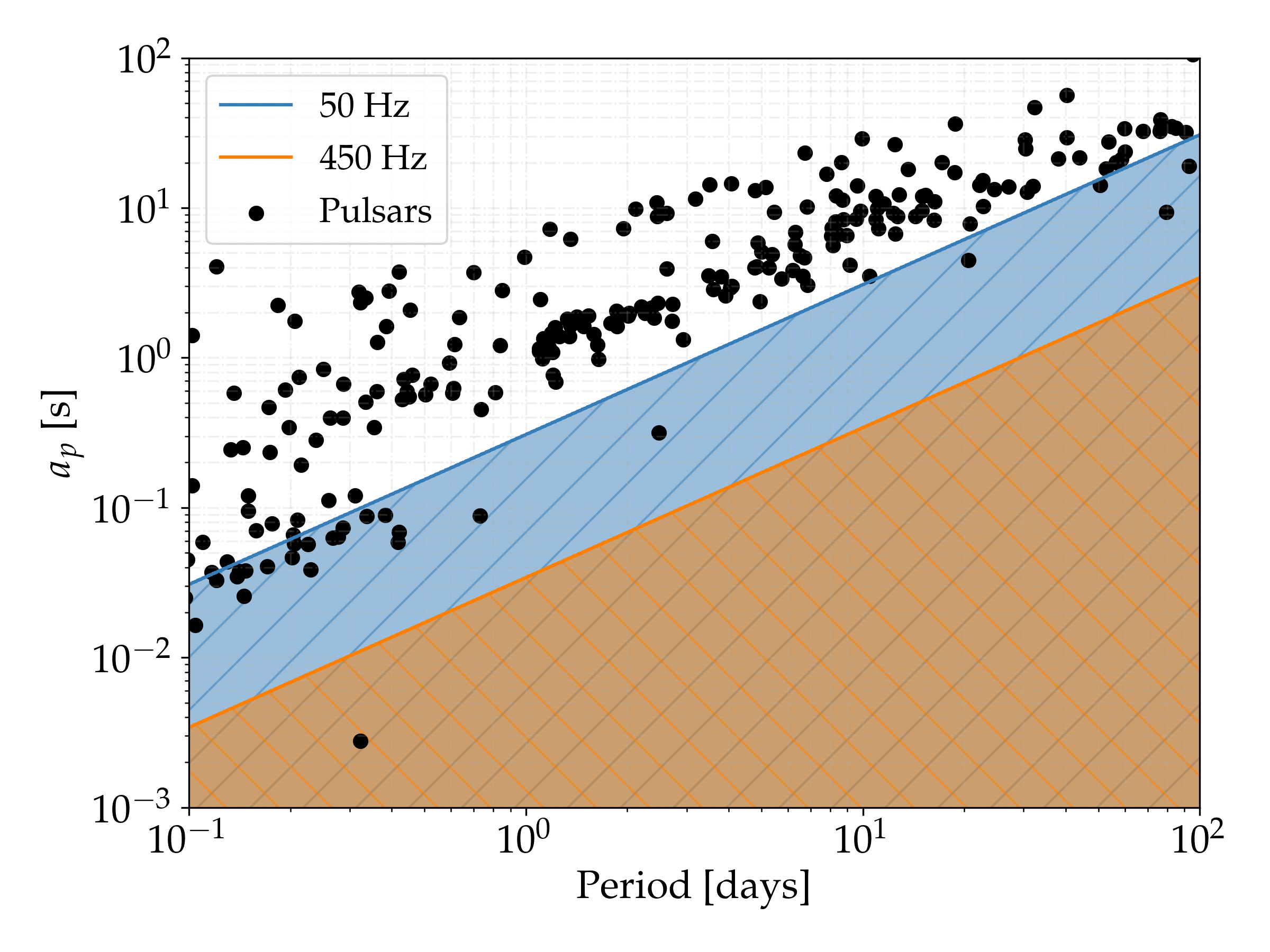}
\caption{The filled areas mark the regions in binary parameter space where the observed binary modulation is less than one frequency bin, for $T_c = 900$ s. Data taken from \cite{ATNF} and downloaded with \cite{psrqpy}.}
\label{fig:mod}
\end{figure}

The range in parameter space to be searched is primarily determined by the astrophysical prior information and by the available computational power:
\begin{itemize}
    \item The frequencies to be searched are determined by the sensitive frequency band of the detectors and by the expected frequency of the emitted gravitational waves. Figure \ref{fig:p-pdot} shows that the maximum gravitational-wave frequency is around 1400 Hz. For the past O1 and O2 observing runs performed by the Advanced LIGO detectors, the most sensitive frequency band ranges roughly from $50$ to $1000$ Hz, with the best strain sensitivity occurring near 150 Hz. %A neutron star with a non-axisymmetric crust is expected to emit gravitational radiation at a frequency twice its rotational frequency; hence stars with spin frequencies of $25 \leq \nu \leq 500$ Hz are of most interest. There are presently 182 known pulsars with observed spin frequencies greater than 25 Hz, and, of these pulsars, 111 are located within binary systems.
    \item The range of binary orbital periods is bounded by the coherent time: periods lower than the coherent time cannot be distinguished one from another, and the equation for the period resolution was derived assuming $T_c \ll P$. The upper bound for the periods to be searched is mostly determined from the astrophysical prior information, where we can see that the maximum period is around $10^3$ days. %and by the observing time (periods higher than it cannot be untangled). %Figure \ref{fig:P0-ap} shows the orbital periods of the detected pulsars in binary systems. 
    %to search depends on several factors, and is described in more detail in section 4. In summary, the number of neutron star orbits during the observation time determines the upper bound of the search over the orbital period. Simulations show that a reasonable upper bound on the orbital period range is one-fifth of the total observation time, Tobs. Meanwhile, the lower bound is governed by the coherence time of the initial Fourier transformation used to cover the parameter space and by the highest frequency to be searched. From this limit, the shortest period is Pmin = 2 h.
    \item The minimum value of $a_p$ is bounded by the minimum Doppler shift that we can observe.  Figure \ref{fig:mod} shows that for a given period, a minimum $a_p$ value needs to be selected in order for the frequency modulation to be higher than 1 bin. If the modulation is less than 1 bin, we are not able to distinguish different templates, and pipelines which look for GW signals from isolated NS can already discover them. The maximum $a_p$ value to be searched is determined by the maximum amount of frequency bins that we can load at the same time, limited by RAM (Random Access Memory), and by the astrophysical prior information extracted from the known pulsar population, which figure \ref{fig:mod} shows.
    \item The range in time of ascending node $t_{\text{asc}}$ that needs to be searched is uniquely determined by the orbital period. Since we can redefine the time of ascension for every orbit by adding an integer times the orbital period, we can define it in the orbit which is closer to the mid-time of the search and we only need to search this area:
    \begin{align}
    t_{mid} - \frac{P}{2} \le t_{\text{asc}} < t_{mid} + \frac{P}{2}.
    \end{align}
\end{itemize}

\subsubsection{Number of templates}
After having defined the spacing between templates and the ranges that have to be covered in each dimension, the total number of templates which are needed to carry out the search can be calculated. 

The scaling of the total number of templates $N$ is given by (and $N_I$ for an all-sky search for isolated NS):
\begin{align}
    dN &\propto T_c^6 T_{obs} f^5 \Omega^4 a_p^2 d\lambda, \\
    dN_I &\propto T_c^4 T_{obs} f^2 d\lambda,
\end{align}
where $d\lambda$ represents a volume element for each of the search dimensions. For the binary case, the scaling with frequency is much steeper than for the isolated case, which will greatly increase the computational cost at higher frequencies. 

\begin{figure}[tbp]
%\begin{center}
%\includegraphics[width=1.0\columnwidth]{NumberTemplatesBinary.png}
\includegraphics[width=1.0\columnwidth]{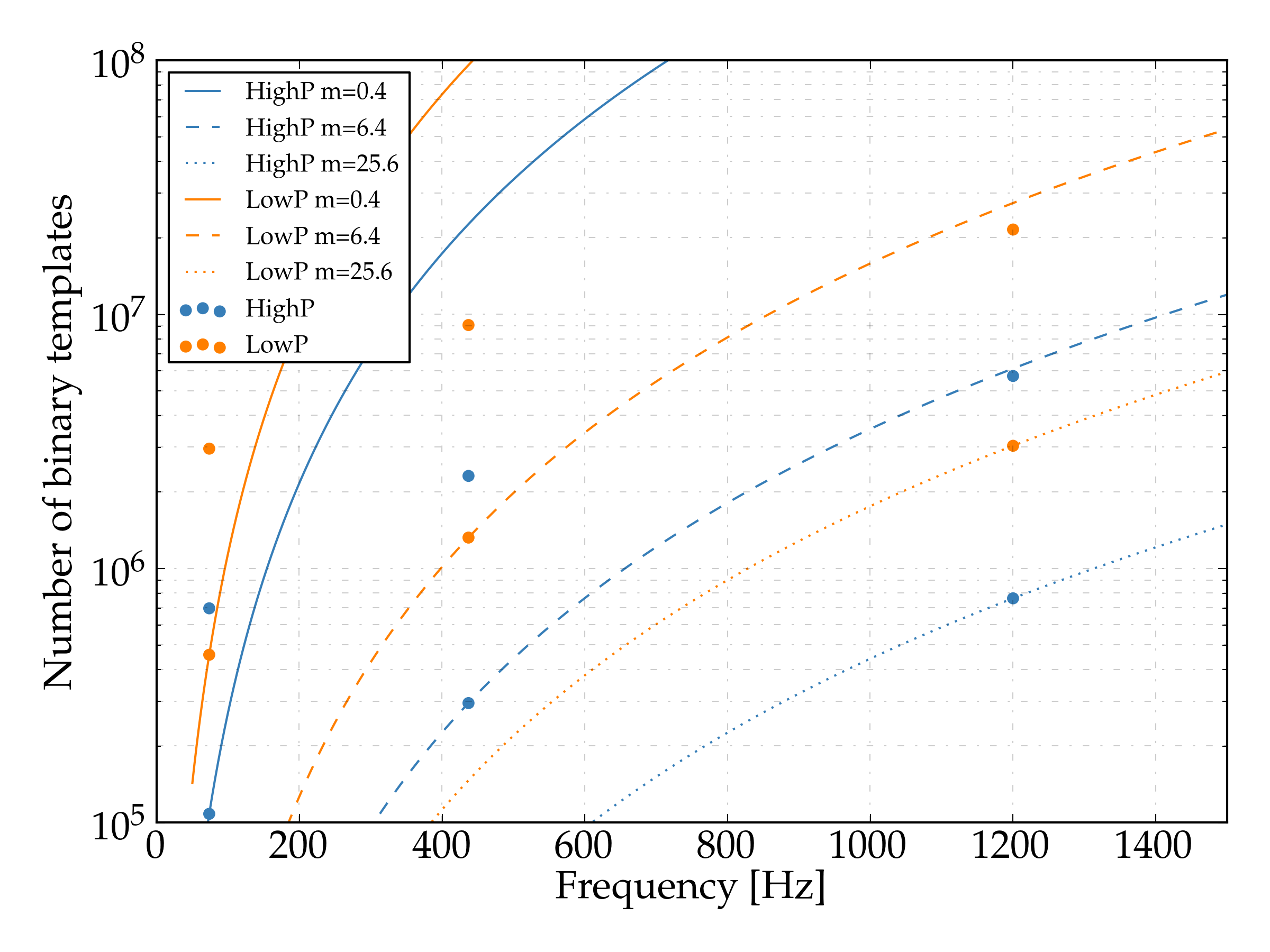}
\caption{\label{fig:binarytempl}Number of binary templates for two searches covering different binary parameter spaces: For ``HighP", $P$ ranges from 10 to 30 days and $a_p$ from 9 to 25 seconds, while for ``LowP" $P$ ranges from 0.1 to 0.15 days and $a_p$ from 0.05 to 0.08 seconds. For each frequency, two different mismatch parameters are chosen (shown with circles): 0.1 and 0.4 for 73.6 Hz; 1.6 and 6.4 for 436.9 Hz; 6.4 and 25.6 for 1200 Hz. The coherent time is 900 s, and the observation time is 11178584 s (the duration of the O1 run).}
%\end{center}
\end{figure}

Figure \ref{fig:binarytempl} shows the number of binary templates which need to be covered for different mismatch configurations and for searches at different regions of binary parameter space. With these numbers, the difference between a search for NS in isolated systems and a search for NS in binary systems becomes clear: while for the former around $\mathcal{O}(10^2)$ spin-down/up templates need to be searched (to cover a range wider than the astrophysical informed range), for the latter more than $\mathcal{O}(10^5)$ templates are needed to cover a narrow astrophysical range. Figure \ref{fig:binarytempl} also clearly shows that if we want to cover a broad range in frequency, the mismatch parameters will have to increase for higher frequencies, because a search with constant low mismatch (like 0.4) is unfeasible otherwise. Another way to solve this issue would be to decrease the coherent time as the frequency increases.

%With limited computational power, a decision has to be done over doing a broader search trying to cover a big binary and frequency parameter space, or a deep search which selects a small portion of the parameter space and has less mismatch, which increases the sensitivity of the search. The optimal way to select the search parameters given some constrain like a maximum computational cost (as discussed in \cite{OptimalSetup}), involves estimating the probability of detection and the computational cost at any region of parameter space and obtaining the best configuration, but this is not in the scope of this paper. By looking at some plots and the scalings of the number of templates, we can make intuitive decisions. For example, the number of binary templates needed to cover a period range between 0.1 and 1 days is bigger than what is needed to cover the range between 1 and 100 days. We could aim to do a broad search and cover everything, but to increase the chances of detection it might be better to focus on the higher periods range, because with the same computational cost a larger range with lower mismatch can be covered. %With the steep scaling with orbital period and frequency,

%With limited computational power, an optimal search strategy would need to use different resolution parameters at different frequencies if a broad range in frequency needs to be covered due to steep scaling with frequency. 

%\subsection{\label{sec:gpu}GPU implementation}

%The BinarySkyHough code is written in C and CUDA and is part of the LSC (LIGO Scientific Collaboration) Algorithms Library (LALapps) repository \cite{LALSuite}.

%After extensively profiling the code, we observed that the majority of the time was spent of one single function, which had a loop over the binary/spin-down templates. This loop was easily parallelizable, since no dependencies between different templates exist. 

%Details about GPU functions

\subsection{\label{subsec:CohTime}Maximum coherent time}

The sensitivity of a semi-coherent method increases with the coherent time, so in principle one should aim to use coherent times as long as possible. On the other side, the computational cost depends on the coherent time, which sets a limit to this value. Furthermore, the spread of power to neighbouring bins (if the frequency of the signal occupies more than one frequency bin during one SFT) limits the maximum SFT time baseline that can be used (which for our method is equal to the coherent time). 

To recover the maximum possible power from the signal, we have to avoid spectral leakage to neighbouring frequency bins. To achieve this, we demand that the signal be contained in half a single frequency bin, which imposes a maximum coherent time:
\begin{align}
    \frac{\Delta f}{2} = \frac{1}{2T_c} \geq \dot{f} T_c \longrightarrow T_c \leq \frac{1}{\sqrt{2|\dot{f}|_{max}}}.
    \label{eqmaxcoh}
\end{align}

\begin{figure}[tbp]
%\begin{center}
\includegraphics[width=1.0\columnwidth]{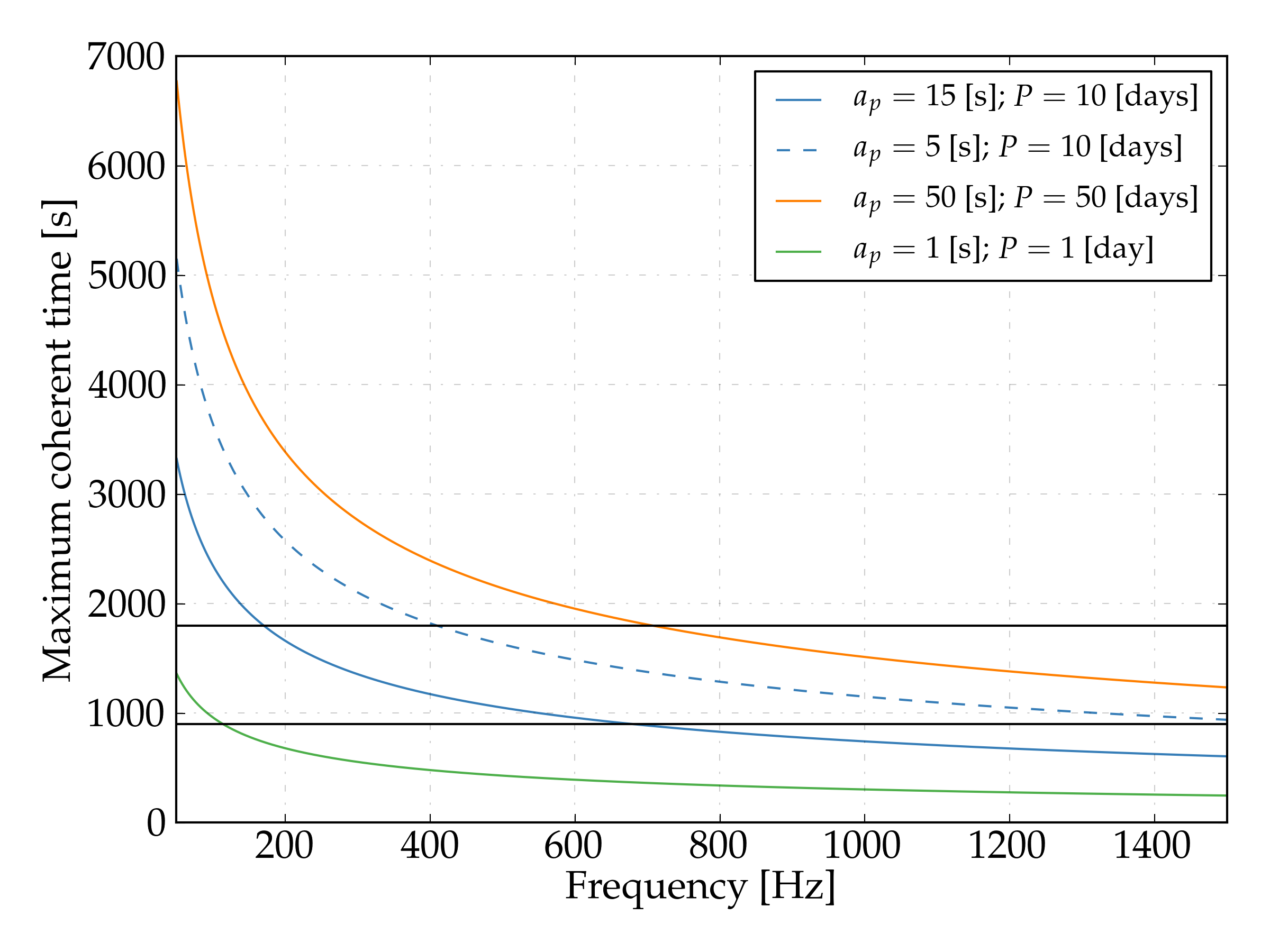}
\caption{\label{fig:coherenttime}Maximum coherent time allowed by equation \eqref{eqmaxcoh} for four different choices of binary parameters. The black horizontal lines mark 1800 s and 900 s.}
%\end{center}
\end{figure}

We can estimate the maximum frequency derivative through the frequency evolution model from equation \eqref{eq:frequencyevoFinal}:
\begin{align}
    \dot{f} = f_0\frac{\vec{a}\cdot\hat{n}}{c} + f_0 a_p \Omega^2 [\sin{\Omega(t-t_{\text{asc}})}],
    \label{eq:derivative}
\end{align}
where $\vec{a}$ is the acceleration vector of the detector in the SSB frame. The highest contribution to the acceleration due to detector motion comes from the daily rotation of the Earth \cite{SkyHough}, which simplifies the previous equation to:
\begin{align}
    |\dot{f}|_{max} = \frac{f_0}{c} \frac{v^2}{R_e} + f_0 a_p \Omega^2 = \frac{f_0}{c} \frac{4\pi^2 R_e}{T_e^2} + f_0 a_p \Omega^2,
\end{align}
where $R_e$ is the radius of the Earth and $T_e$ is a sidereal day. It can be seen that the maximum coherent time depends on the frequency $f_0$ and on the binary parameters $a_p$ and $\Omega$. Figure \ref{fig:coherenttime} shows some examples of these dependencies. It can be seen that lower coherent times are able to cover wider ranges of parameter space.

The optimal search strategy should use SFTs with different coherent time (the maximum allowed) in different regions of the binary and frequency parameter space. We note that the curves shown in figure \ref{fig:coherenttime} are conservative, since they are assuming the worst case scenario, and usually we will be in a more relaxed case where we could use longer coherent times. These assumptions are that the two frequency derivative terms are at their maximum values and they are aligned, and that we only allow half a frequency bin of variation, because we assume that the frequency of the signal is at the center of the bin, which almost never happens.

If the eccentricity is non-zero, two more factors would be present in equation \eqref{eq:derivative}. For eccentricities smaller than 0.01, it can be seen that these factors (proportional to $e$) would contribute much less to the frequency derivative. For this reason, we don't take them into account in our previous estimation.

\subsection{Maximum eccentricity}
\label{subsec:MaxEcc}

Our final model for the frequency evolution, given by equation \eqref{eq:frequencyevoFinal}, assumes a zero-eccentricity orbit, but our pipeline remains fully sensitive to signals with a certain eccentricity if this eccentricity does not produce a noticeable change (less than a frequency bin) in the frequency-time evolution.

We can estimate the error in frequency tracking  (by assuming that the eccentricity is exactly zero when it is non-zero) by subtracting equations \eqref{eq:frequencyevo} and \eqref{eq:frequencyevoFinal}:
\begin{align}
    |\Delta f(t)| = 2 f_0 a_p \Omega [ &\frac{\epsilon \cos{\omega}}{2} \cos{[2\Omega(t-t_{\text{asc}})]} \nonumber \\
    + &\frac{ \epsilon \sin{\omega} }{2} \sin{[2\Omega(t-t_{\text{asc}})]} ].
    \label{eq:frequencyevoerror}
\end{align}
The maximum error at any time will be:
\begin{align}
| \Delta f |_{max} &= \epsilon f_0 a_p \Omega.
\end{align}

We demand that this frequency difference is smaller than half a frequency bin, as was done in the previous subsection:
\begin{align}
    \epsilon f_0 a_p \Omega \leq \frac{1}{2 T_c} \longrightarrow \epsilon \leq \frac{1}{2 T_c f_0 a_p \Omega}.
    \label{eq:frequencyevoerror2}
\end{align}

%If $e$ is less than or equal to this, the change that is produced by not taking it into account does not change the estimated frequency bin.
If $e$ is less than or equal to this, the calculated frequency evolution will not deviate by more than half a frequency bin from the true evolution. Again, this expression depends on the region of the binary and frequency parameter space that we are in, so searches at different parts of this space can remain sensitive to different levels of eccentricity.

\begin{figure}[tbp]
%\begin{center}
\includegraphics[width=1.0\columnwidth]{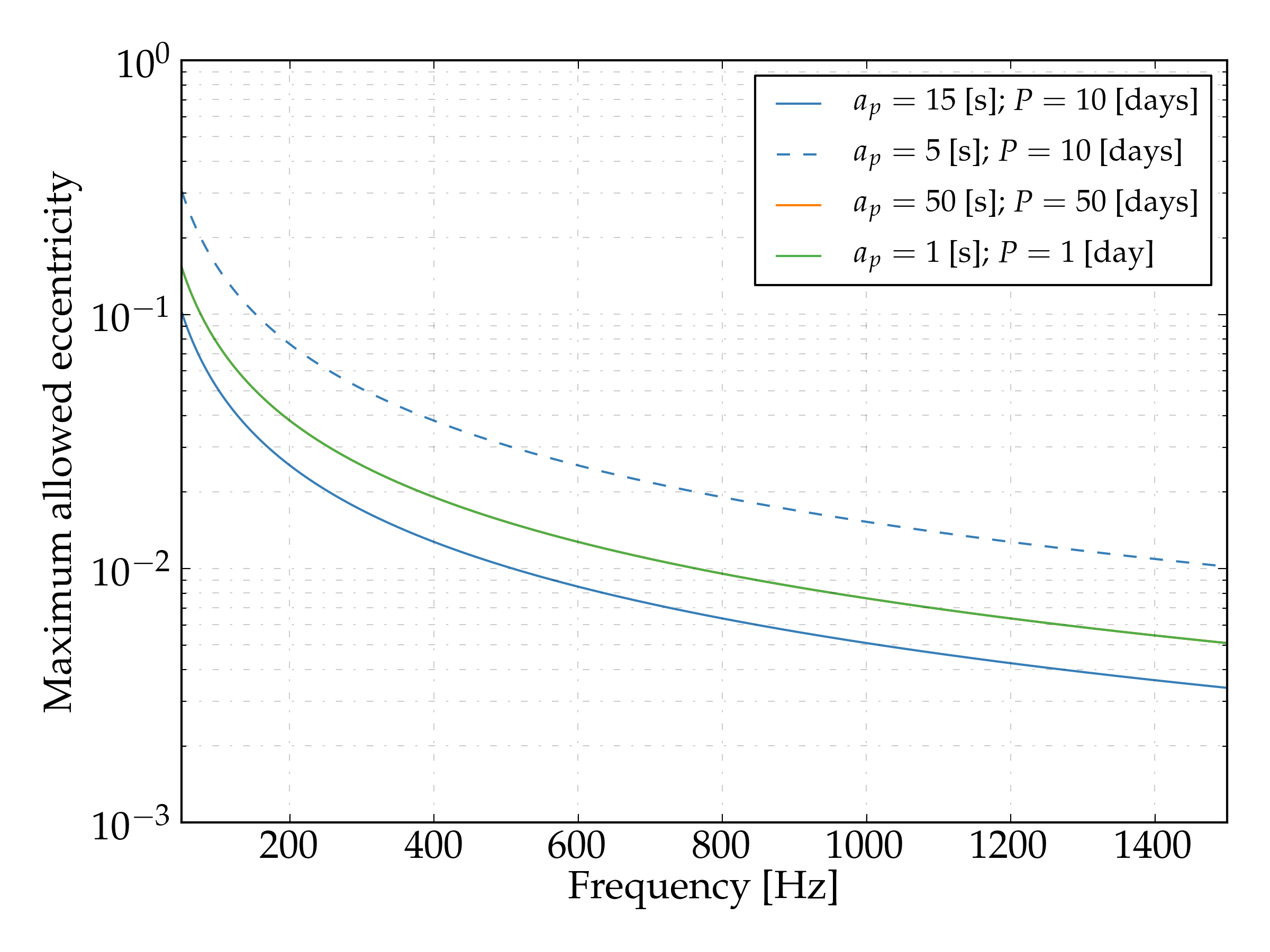}
\caption{\label{fig:maxeps}Maximum eccentricity allowed by equation \eqref{eq:frequencyevoerror2} for three different choices of binary parameters, with a constant choice of $T_c = 900$ s. The third and fourth traces overlap.}
%\end{center}
\end{figure}

For most of the SFTs this error will be smaller (again, this is a conservative estimation), so this is a lower limit on the eccentricity that the orbit of the neutron star can have without producing any noticeable difference. Figure \ref{fig:maxeps} shows that the zero-eccentricity assumption does not affect our ability to track systems with eccentricity smaller than 0.01 (for frequencies lower than 500 Hz for the worst case shown in that figure). In section \ref{sec:sensitivity} we will show some simulations of how the eccentricity affects the sensitivity of our method. An estimation of the sensitivity lost for signals with eccentricities higher than this is left for future work.

%\subsection{Relativistic corrections}
%\label{subsec:RelEff}

%Post-Keplerian parameters give relativistic corrections to the classical Keplerian orbits described in section \ref{sec:sigmodel}. With our search method, one of the five PK parameters that we could observe is the decay in orbital period, which for zero-eccentricity orbits is given by:
%\begin{align}
%    \dot{P}_b = \frac{-192 \pi G M_{\odot}}{5c^3} \frac{M_C M_{NS}}{(M_C + M_{NS})^{1/3}} (\frac{2\pi}{P_b})^{5/3}.
%    \label{eq:periodder}
%\end{align}

%Other relativistic effects, such as geodetic precession, also have a negligible effect in our analysis.

\subsection{\label{sec:postpro}Post-processing}

%The post-processing explained here is very similar to the one that has been applied in other searches like \cite{O1CWAllSkyFull}. We will briefly summarize it and explain the differences.

After finishing the main steps of the pipeline, we are left with one top-list per dataset for each region in parameter space (i.e. each 0.1 Hz band), which contains the top templates ordered by a detection statistic. If the pipeline was run with option A, we will have multiple top-lists, while if it was run with option B a single top-list will be the output. The following steps detail the procedure which goes from this point to the final list of outliers:

\begin{enumerate}
    \item If we have multiple top-lists, the first step consists of searching for coincidental pairs between these lists, by calculating the distance in parameter space and selecting the pairs which are closer than a certain threshold. The optimal value for the threshold cannot be found analytically, since its value depends on a balance between being more sensitive and having too many outliers. A reasonable value can be found by doing simulations. For each coincidental pair, its centroid (average locations in parameter space weighted by significance) is calculated. The distance $d$ in parameter space is given by:
    \begin{align}
    d^2 = &\left(\frac{\Delta f}{\delta f}\right)^2 + \left(\frac{\Delta x}{\delta \theta}\right)^2 + \left(\frac{\Delta y}{\delta \theta}\right)^2 \nonumber \\ + &\left(\frac{\Delta a_p}{\delta a_p}\right)^2 + \left(\frac{\Delta \Omega}{\delta \Omega}\right)^2 + \left(\frac{\Delta t_{\text{asc}}}{\delta t_{\text{asc}}}\right)^2,
    \label{eq:postdist}
    \end{align}
    where the numbers in the numerators represent the difference between two templates and the numbers in the denominators represent the parameter resolution. This distance is dimensionless and is given as a number of bins. The quantities $x$ and $y$ represent the cartesian ecliptical coordinates projected into the ecliptical plane.  

    When we compare two templates which are at different parts of the parameters space, the resolution of the binary parameters is different at these points. To calculate the distances of equation \eqref{eq:postdist}, we use a mean of the resolution at both points. 
    
    \item Independently of running the search with option A or option B, now we have a unique list. The next step consists of searching for clusters in this list. This will group different templates which can be ascribed to the same cause, and will reduce the number of final outliers. Again, we set a threshold in parameter space distance and find candidates which are closer than this distance. Clusters are found by analyzing the distance of each template from all other templates, and keeping a list of indices of members with distances below the threshold. Each template can only be part of a cluster, so if a template was already in a cluster its newly generated cluster and the old one will be joined to form a unique cluster.
    
    If the search is run with option B, another distance threshold between the top template in a cluster and all the other members is calculated, which eliminates all members which are further away than a certain threshold. This step is not needed with option A since the coincidence step already eliminates many candidates, but with option B the clusters can grow too wide and the final parameter estimation can be wrong unless a cut is made, due to a high number of cluster members being too distant form the true signal values.
    
    \item The final post-processing step consists on calculating the centroid of each cluster. This is calculated as a weighted (by significance) average among all the members of the cluster. We keep the most significant cluster per 0.1 Hz band (if any), selected by the summed significance of all its members. This produces the final list of outliers of the search.
    
    We cannot claim that these outliers represent a real astrophysical signal. As known from previous searches, instrumental noise or spurious coincidences can end up in the final list. As shown in figure \ref{fig:flowchart}, the final steps of any CW pipeline are the application of vetoes and follow-up procedures which increase the significance of the candidates and enhance the parameter estimation. Due to the extra parameters needed for a search for NS in binary systems, follow-up procedures used in past searches may not be applied to this case. A derivation of a follow-up procedure which can increase the confidence on candidates from the \textit{BinarySkyHough} pipeline is left for future work. %For this reason, vetos which demand that these outliers have certain properties, and follow-up procedures which increment the significance of these outliers are used in order to increment the confidence of a detection or to remove the non-astrophysical remaining outliers.
\end{enumerate}

%The main difference between the procedure used in previous searches and the one needed for this algorithm is that the parameter space distance has to take into account the binary parameters. This distance now is:

\subsection{\label{sec:compcost}Computational model}

The sensitivity of CW searches is always limited by the available computational budget. Usually, a choice must be made between doing a broader search trying to cover a large binary and frequency parameter space, or a deep search which selects a small portion of the parameter space and has less mismatch, which increases the sensitivity of the search. In order to estimate \textit{a priori} the cost that a search will have and to compare different setups, it is important to construct a computational model which can estimate the total computational cost and required RAM of a search given some regions of parameter space and resolution parameters. %The optimal way to select the search parameters given some constrain like a maximum computational cost (as discussed in \cite{OptimalSetup}), involves estimating the probability of detection and the computational cost at any region of parameter space and obtaining the best configuration, but this is not in the scope of this paper. By looking at some plots and the scalings of the number of templates, we can make intuitive decisions. For example, the number of binary templates needed to cover a period range between 0.1 and 1 days is bigger than what is needed to cover the range between 1 and 100 days. We could aim to do a broad search and cover everything, but to increase the chances of detection it might be better to focus on the higher periods range, because with the same computational cost a larger range with lower mismatch can be covered.

%As mentioned before, with a limited computational budget a balance between doing a broad search (spanning a larger region of parameter space) or a deep search (minimizing the mismatch parameters) has to be found. %With such model we are able to prioritize what we want to do and to better estimate where the resources can be better spent.

\subsubsection{Computational cost}

The code spends the majority of the time in the inner-most loop over SFTs, sky positions and frequency and binary templates that is done at the first step of the search, and a loop over the SFTs and a subset $N_C$ of the total number of templates at the second step of the search. To estimate the total cost of a search, we will characterize the scaling of these parts of the code by running it with different search parameters.

The total cost will be slightly bigger than this estimation, due to other tasks like I/O and initialization of variables, but this extra cost is negligible. Furthermore, we are going to estimate the cost for frequency bands which are nearly Gaussian (i.e. don't contain signals or excessive instrumental noise), and we will assume that this will be valid for the vast majority of analyzed frequency bands. %The computational cost produced by the second step of the search (where the power significance detection statistic is calculated for the top templates is calculated) is many orders of magnitude smaller than the cost of the first step, and for this reason we don't estimate it.%, since the vast majority of bands and sky-patches are of this kind. This will facilitate our cost estimation.

The computational cost can be estimated as:
\begin{align}
    \mathcal{C} = N_B N_D N_F \sum_{i=0}^N N_{SP;i} (C_{1;i} + C_{2;i}),
    \label{eq:cost}
\end{align}
where $N_B$ is the number of blocks of binary templates that must be analyzed to cover the entire range of binary parameters, $N_D$ is the number of datasets (assuming they have the same number of SFTs or taking the maximum between all of them), $N_F$ is the number of frequency templates in each frequency band (the same number for each frequency band, e.g. 90 for a 0.1 Hz band with $T_c=900$ s), and the summation goes over the number of frequency bands in which we split the total frequency range. To cover a wide range of binary parameters, more than one block of binary templates is usually needed, since the maximum number of binary templates that can be searched at once is limited by RAM constraints, as will be seen in subsection \ref{sec:RAM}. %Furthermore, as can be seen in figure \ref{fig:mod}, at different periods the range in $a_p$ to be searched varies.

The number of sky-patches in the \textit{i}th frequency band $N_{SP;i}$ can be estimated as:
\begin{align}
    N_{SP;i} \approx \frac{4\pi}{N_X N_Y \delta \Theta^2} = \frac{4\pi (v/c)^2 P_F^2 T_C^2 f_i^2}{N_X N_Y},
\end{align}
being $N_X$ and $N_Y$ the number of sky pixels in each direction.

The cost $C_{1;i}$ of the first step of the search per one sky-patch and one frequency template is given by:
\begin{align}
    C_{1;i} = C_{B1} N_T N_{SFTs} N_X N_Y \rho_F,
    \label{eq:costT}
\end{align}
where $N_{SFTs}$ is the number of SFTs, $N_T$ is the number of binary templates in one binary block at the \textit{i}th frequency band, $\rho_F$ is a factor which controls the scaling with the selected peak threshold and $C_{B1}$ is the cost of running over 1 binary template when there is one SFT and one sky position. $C_{B1}$ is calculated when the threshold is $\rho_{th}=1.6$, and $\rho_F = \frac{e^{-\rho_{th}}}{e^{-1.6}}$ is a simple scaling factor which takes into account the different number of peaks that are present (the exponentials appear since the distribution of powers for a Gaussian band is $p(\rho_k)=e^{-\rho_k}$) when the threshold is changed.

The cost $C_{2;i}$ of the second step of the search per one sky-patch and one frequency template is given by:
\begin{align}
    C_{2;i} = C_{B2} N_C \mathcal{N} \bar{N}_{SFTs},
    \label{eq:cost2}
\end{align}
where $\mathcal{N}$ is the total number of templates at the \textit{i}th frequency band, $\bar{N}_{SFTs}$ is the number of SFTS used at the second step and $C_{B2}$ is the cost of the second step per template and per SFT.

%Due to the usage of GPUs and the way the loops are arranged, we have found non-linearities in equation \eqref{eq:costT}. For example, doubling the number of SFTs may not double the timing of a run, and the same can happen when changing $S_i$ (the number of sky positions). For this reason, we join some of these factors and our timing equation now reads:
%\begin{align}
%    C_i = C'_{B}.
%    \label{eq:costT}
%\end{align}

The code is run with a CPU and GPU. The code for the GPU execution is written in CUDA, which has some parameters that can be changed (like the number of blocks and the number of threads per block) which affect these estimations. Another important source of uncertainty is the different hardware layouts between different GPUs, like the different number of cores. These differences do not affect the predicted scalings given by the previous equations.%, and only enter the previous equation through $C_B$.

We have done several runs to test the different scalings. We have used two different GPUs, and we also show a comparison by using only a CPU instead of a CPU+GPU. The results are shown in table \ref{tab:SkyHoughCompCost}. The listed configurations on blocks and threads for the GPUs are the ones that have given better results. It can be seen that without using a GPU card this search would be unfeasible. %The analyzed binary space is the same ``HighP" as in figure \ref{fig:binarytempl}, an we analyze a 0.1 Hz frequency band.

From these results we can estimate the cost that a complete search would have. For 1 binary block of $3\times10^{5}$ binary templates, with 1 dataset, 90 frequency bins per 0.1 Hz covering 400 Hz, and 50 sky-patches per 0.1 Hz band, a search with configuration run 2 would need approximately $5000$ hours. This assumes that the number of binary templates would be the same in each frequency band, which requires that the mismatch parameters are lowered as the frequency is increased.%, and that the number points per sky-patch remains constant, which makes the number of sky-patches increase with frequency.% (the total number of sky-patches fro a frequency range from 50 to 450 Hz is X).

If we want the binary resolution to remain constant across the frequency range, the number of binary blocks would inrease with frequency. This number would also be greater than 1 if we wanted to cover a larger range of binary parameters. With the same configuration as before but with 500 binary blocks, the cost would increase to $2.5\times 10^6$ hours. This order of magnitude is usual within all-sky semi-coherent searches, and is comparable to the cost of the only published all-sky search \cite{TwoSpectResults}. Even though this method explicitly searches over $t_{\text{asc}}$, which \textit{TwoSpect} does not, the costs are comparable due to the usage of GPUs and the look-up table approach. With 500 binary blocks we could cover a large parameter space, covering all the astrophysical interesting regions, or we could do a narrower search with low mismatch.

\begin{table*}[tbp]
\begin{center}
\begin{tabular}{ c c c c c c c c c c c c c}
\hline
 & \multicolumn{2}{c}{Run 1} & \multicolumn{2}{c}{Run 2} & \multicolumn{2}{c}{Run 3} & \multicolumn{2}{c}{Run 4} & \multicolumn{2}{c}{Run 5} & \multicolumn{2}{c}{Run 6} \\
Hardware & $C_{1;i}$ & $C_{2;i}$ & $C_{1;i}$ & $C_{2;i}$ & $C_{1;i}$ & $C_{2;i}$ & $C_{1;i}$ & $C_{2;i}$ & $C_{1;i}$ & $C_{2;i}$ & $C_{1;i}$ & $C_{2;i}$ \\
\hline \hline
CPU + Tesla V100 & 0.15 & 0.11 & 0.41 & 0.30 & 0.80 & 0.54 & 0.45 & 0.35 & 2.2 & 1.8 & 0.07 & 0.05 \\
CPU + GTX 1050Ti & 1.7 & 2.1 & 4.5 & 5.6 & 8.5 & 11.0 & 6.6 & 7.8 & -- & -- & 0.8 & 1.1 \\
CPU & 122 & 316 & -- & -- & -- & -- & -- & -- & -- & -- & -- & --\\
\hline
\end{tabular}
\caption{$C_{1;i}$ and $C_{2;i}$ timings (in seconds) for different run configurations. Each number is the mean over 500 runs with the same configuration. Both GPUs are used with 512 blocks and 64 threads per block. The CPU used is an Intel(R) Xeon(R) CPU E5-2650 v4 @ 2.20GHz. The compilation of the code was done with gcc and nvcc, with option -O3. All runs use $N_C=0.05$. \\ Run 1: $N_T=1 \times 10^5$; $N_X N_Y=49$; $N_{SFTs}=13304$; $\bar{N}_{SFTs}=25930$. Run 2: $N_T=3 \times 10^5$; $N_X N_Y=49$; $N_{SFTs}=13304$; $\bar{N}_{SFTs}=25930$. Run 3: $N_T=6 \times 10^5$; $N_X N_Y=49$; $N_{SFTs}=13304$; $\bar{N}_{SFTs}=25930$. Run 4: $N_T=1 \times 10^5$; $N_X N_Y=169$; $N_{SFTs}=13304$; $\bar{N}_{SFTs}=25930$. Run 5: $N_T=6 \times 10^5$; $N_X N_Y=169$; $N_{SFTs}=13304$; $\bar{N}_{SFTs}=25930$.  Run 6: $N_T=1 \times 10^5$; $N_X N_Y=49$; $N_{SFTs}=7235$; $\bar{N}_{SFTs}=14220$.}
\label{tab:SkyHoughCompCost}
\end{center}
\end{table*}

\subsubsection{Random Access Memory (RAM)}
\label{sec:RAM}
In order to characterize the RAM required by our pipeline, a calculation of the number of bytes taken by every data structure should be made. Of the many data structures in the code, two of them are many orders of magnitude larger than the rest and are enough to give a rough estimate of the memory required. 

One of these structures is related to the PHMDs, and has a size in bytes of:
\begin{align}
    S_a = 6 N_{SFTs} K N_X N_Y,
    \label{eq:RAM1}
\end{align}
where $K$ is the number of PHMDs needed in the frequency axis (see figure \ref{fig:PHMD}), equal to the number of searched frequency bins plus the maximum modulation produced by the BB Doppler modulation.

The other large structure holds the results of the first step of the search, and the size in bytes is given by:
\begin{align}
    S_b = 8 N_T N_X N_Y.
    \label{eq:RAM1}
\end{align}
%The other structures have a negligible cost compared to these ones.

%Another important limiting structure in our code are the Hough maps, calculated at the first step of the search, which are stored in GPU shared memory. The maximum size of shared memory limits the number of sky-points in a sky-patch. For the Tesla V100, this limits the number of points to 164.

With these expressions and a number of $N_{SFTs}$ to be analyzed, we can calculate the RAM for different number of binary templates and different number of PHMD bins given by different frequency band sizes. %These will limit the sizes of our jobs or the number of simultaneous CPU processes which can run using a unique GPU.

\section{\label{sec:sensitivity}Sensitivity estimation}

%In this section we present an estimation of the sensitivity of the \textit{BinarySkyHough} pipeline. 

This section presents a characterization of the sensitivity of the \textit{BinarySkyHough} pipeline. To do this, we add many simulated signals to real or simulated noise in a Monte-Carlo way and we run the pipeline with this data as input. We determine the number of detected signals, and we evaluate the parameter estimation obtained. Simulations are used because an analytical estimation of the sensitivity of a pipeline which takes into account all the steps of the procedure cannot usually be obtained. This is a widely used procedure and it has been used in many past searches such as \cite{O2KnownPulsars} or \cite{O1CWAllSkyFull}.

The purpose of this section is twofold: to estimate the sensitivity of the pipeline, and to see how it changes by varying different internal parameters such as the mismatch or the fraction of templates $N_C$ which go to the second step of the search.

%These simulated signals are added at different frequency bands and different parts of the binary parameter space,  %Many methods (like \cite{SkyHough}) develop an analytic estimation of the sensitivity through statistical arguments like false alarm and false dismissal (in a Neyman-Pearson sense), but w

\subsection{Procedure}
We have added signals (usually called injections) into the O1 Advanced LIGO data of detectors H1 and L1 using the commonly used LALSuite code \textit{lalapps\_Makefakedata\_v5} \cite{LALSuite}. We have used 3 different 0.1 Hz bands: 73.6, 170.2 and 436.9 Hz (the signals have a random frequency within each 0.1 Hz) and four different parts of binary parameter space, indicated in table \ref{tab:SkyHoughBin}. We use different levels of amplitude $h_0$, selected to have certain sensitivity depths:
\begin{align}
    \mathcal{D} = \frac{\sqrt{S_n}}{h_0}.
\end{align}
We have used 4 sensitivity depths (14, 18, 22 and 26 Hz$^{-1/2}$) in order to be close to the $95\%$ efficiency point, a percentage usually used to ascertain the sensitivity of a search method, with 100 different signals per sensitivity depth. Other amplitude parameters like cosine of inclination, initial phase and polarisation are drawn from a uniform distribution (producing signals with random polarizations). We have used a coherent time of 900 s for all the studies presented here. The injected signals are isotropically distributed in the sky, with random argument of periapsis $\omega$ and with eccentricity drawn from a log-uniform distribution between $10^{-6}$ and $10^{-2}$. %The effective observation time is equal to 900 times number of SFTs.

In a real search the number of templates which get into the final top-list is limited, and this sets an artificial threshold on the significance of templates which can be detected (if a signal produces a detection statistic with a value lower than this threshold, it won't be present in the final top-list). Before analyzing the injections (which are analyzed in a reduced region around its real parameters, of around 20 bins in each dimension), we run an all-sky search without added signals with the same configuration parameters (parameter space resolution, $N_C$, ...) to obtain this threshold, and we apply it when we analyze the injections, thus ensuring a fair and realistic analysis. The number of candidates per injection that we keep in the final top-list is 5000, the same number that is used for obtaining the threshold in detection statistic.

For each group of 100 signals at each sensitivity depth, we calculate the efficiency, defined as the number of detected signals divided by the number of injected signals, which will be the main indicator of the method's sensitivity. To count an injection as detected, we demand that its final parameters estimated from the selected cluster are within 13 bins of the true parameters, a number which has been used in past analyses (as will be shown later, most injections are recovered at less than two bins away). %We follow the same procedure as in the all-sky search: run the main search and then apply coincidences, clustering and population veto. Translate from sensitivity depth to amplitude using the sensitivity depth given by:

All the results shown in this section use a coincidence window of 3 bins and a clustering window of $\sqrt{14}$ bins. These sizes are similar to the ones that were used for the isolated-star O1 and O2 analysis, and we leave for future work a proper characterization of the effect that these sizes have on the sensitivity and parameter estimation. All the injections have been analyzed with a threshold in power of $\rho_{th}=1.6$. 

\begin{table}[tbp]
\begin{center}
\begin{tabular}{ c c c c}
\hline
Name & $a_p$ [s] & Period [days] \\
\hline \hline
BS1 & 0.03 -- 0.08 & 0.1 -- 0.101 \\ 
BS2 & 0.5 -- 1.5 & 1 -- 1.01\\ 
BS3 & 3 -- 13 & 10 -- 20 \\
BS4 & 20 -- 35 & 30 -- 90 \\
\hline
\end{tabular}
\caption{Four ranges of binary parameters used for the simulations.}
\label{tab:SkyHoughBin}
\end{center}
\end{table}

Before discussing the results, we want to remark that the efficiency or the $95\%$ sensitivity depth are not the unique indicators of the value of a pipeline. Other factors, such as the range in parameter space which can be covered (the computational cost), the robustness to deviations of the signal from the model or to noise artifacts from the detectors, or the parameter estimation are also important indicators. %Furthermore, the sensitivity depth can only be used to compare different pipelines using the same dataset, because it does not take into account other factors such as the observation time which also play an important role.

\subsection{Results}

We have analyzed the simulations by running the pipeline with varying parameters, such as mismatch, and we compare the results obtained in order to get a general view of the sensitivity which this pipeline can achieve. The plots are shown without error bars in order to ease viewing the results, but all these efficiency points should have a vertical error bar equal to $\sqrt{\frac{E(1-E)}{100}}$, where $E$ is the efficiency.

\begin{figure}[tbp]
\includegraphics[width=1.0\columnwidth]{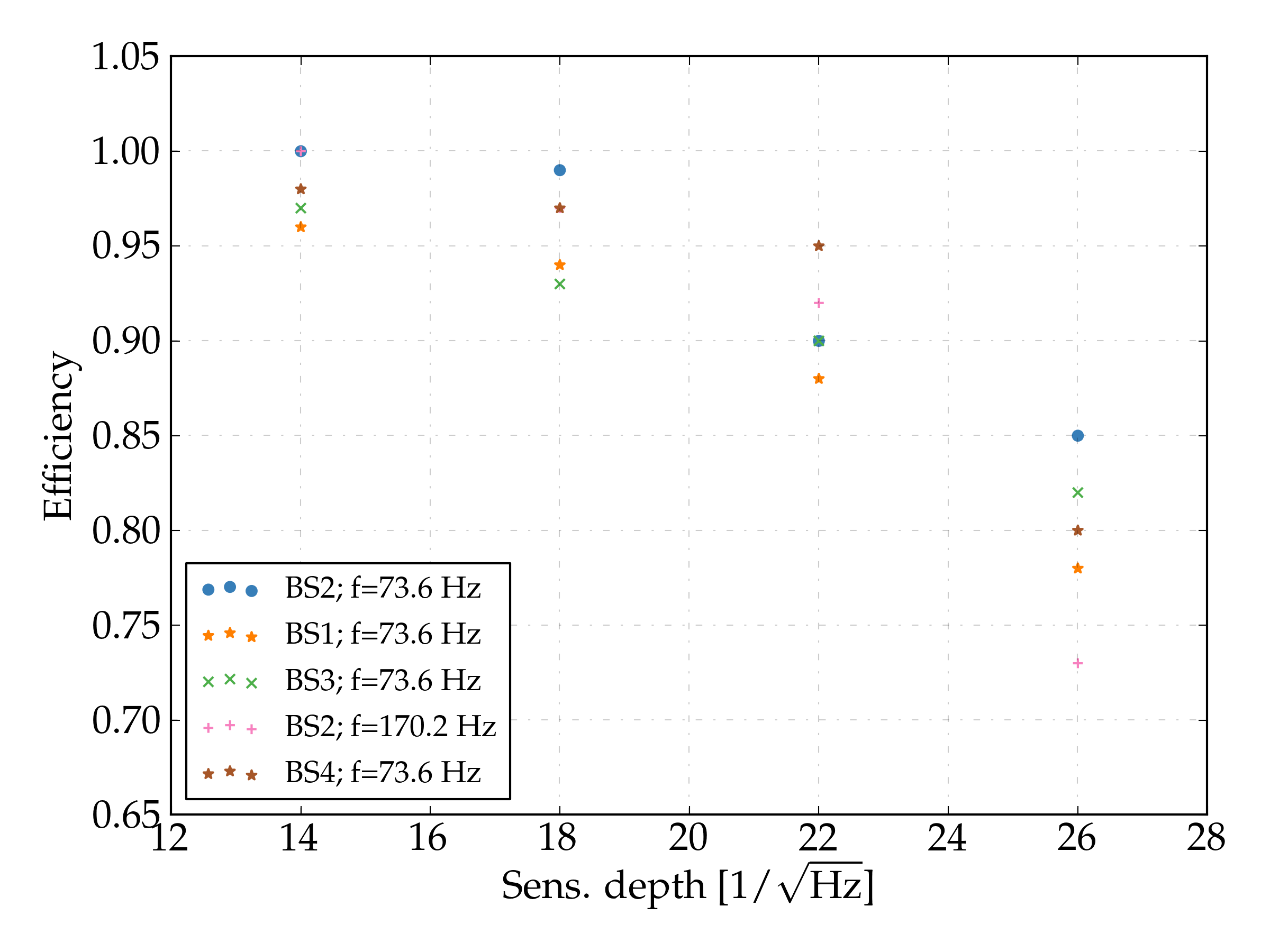}
\caption{Efficiency versus sensitivity depth at different parts of the binary parameter space and two different frequencies. All injections have been analyzed with option B (H1+L1 data), $N_C=0.05$, $m_x=0.4$ and $P_F=1$.}
\label{fig:sensitivity1}
\end{figure}

\begin{figure}[tbp]
\includegraphics[width=1.0\columnwidth]{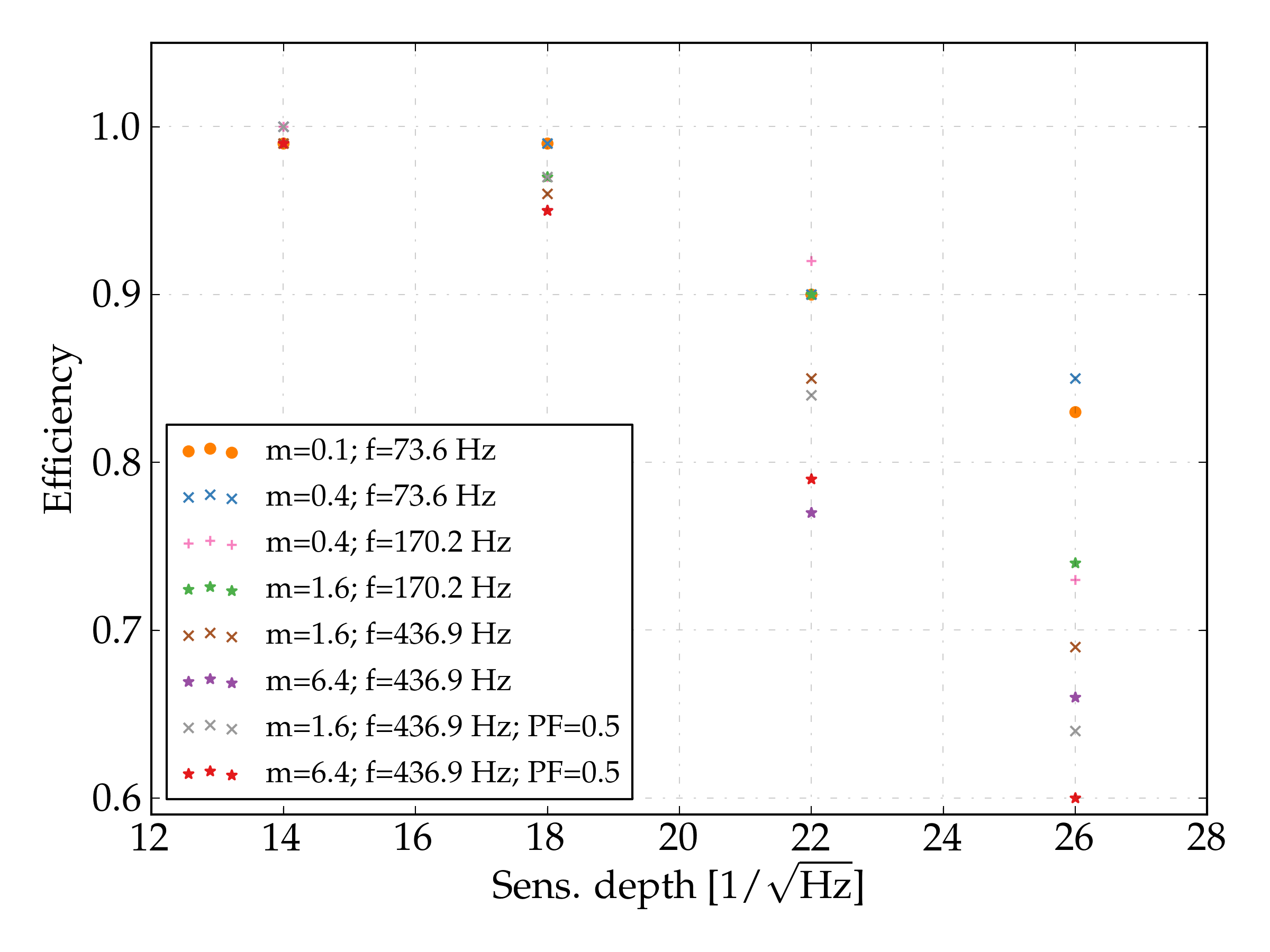}
\caption{Efficiency versus sensitivity depth for different mismatch configurations. All injections have been analyzed at the binary space 2, with option B (H1+L1 data) and $N_C=0.05$.}
\label{fig:sensitivity2}
\end{figure}

\begin{figure}[tbp]
\includegraphics[width=1.0\columnwidth]{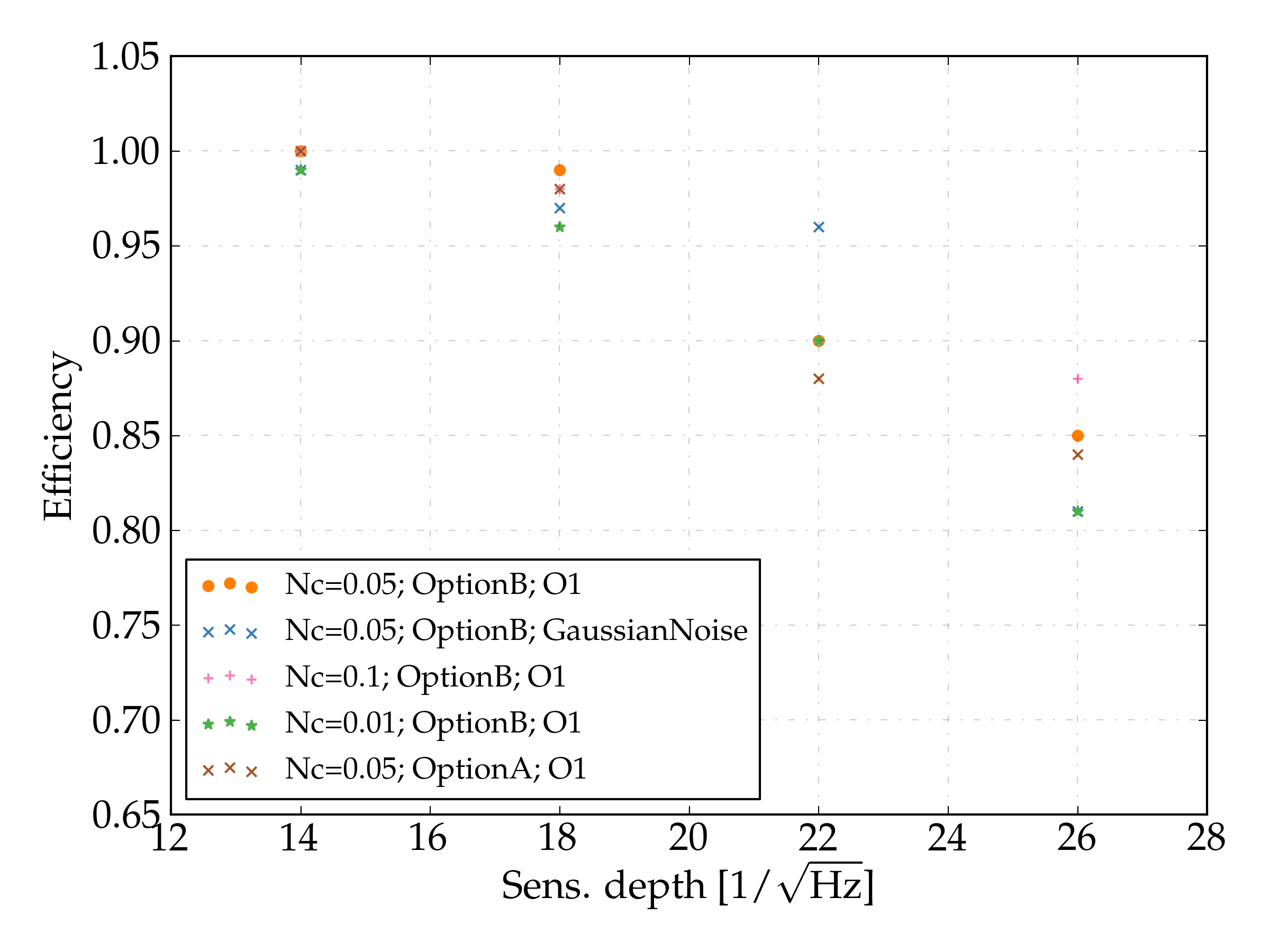}
\caption{Efficiency versus sensitivity depth for different run configurations. $N_C$ shows the fraction of total templates which are passed to the second step of the search. All injections have been analyzed at the binary space 2, with $m_x=0.4$ and $P_F=1$.}
\label{fig:sensitivity3}
\end{figure}

Firstly, figure \ref{fig:sensitivity1} shows a comparison between different parts of binary space and two different frequencies. These four runs share the same mismatch parameter of $m=0.4$. All these choices of binary parameters have approximately the same number of templates, and they all produce frequency modulations wider than one frequency bin. We can observe that all of them have a $95\%$ sensitivity depth above 14 Hz$^{-1/2}$. For the first three sensitivity depth points, all efficiencies are comparable. We can begin to see a wider spread at the last point, where the injections at a higher frequency band show the worse sensitivity.

Secondly, figure \ref{fig:sensitivity2} shows a comparison of runs with different resolution parameters, for the binary space 2. For the first and second sensitivity depth points all efficiencies are very similar. A noticeable decrease in efficiency for coarser resolutions only begins to take place at the last two sensitivity depth points. Running with coarser resolutions also affects the parameter estimations results, effect that we later discuss. The results for the other binary spaces have also been obtained and show similar scalings to the ones shown in this figure.

Lastly, figure \ref{fig:sensitivity3} shows the efficiency obtained by comparing runs with different $N_C$ and comparing option A with option B. The figure shows that running the pipeline with option B (all datasets together) gives a better efficiency than option A. We also observe an increase in efficiency when increasing the fraction of templates which go to the second step of the pipeline, but the improvement is not significative until the last sensitivity depth point. %, and the benefits of running with this number are not compensated by the extra computational cost. 
A comparison with using Gaussian data (of the same noise level) instead of O1 data is also shown. The results for Gaussian noise have been obtained by averaging over 10 different realizations of Gausssian noise. It can be seen that for these Gaussian noise realizations, results are only greatly improved at the third sensitivity depth point, with no significant improvements at the other points. These results have been also obtained for other mismatch configurations and other regions of binary parameter space and they show similar behaviour. %We also compare the efficiency with different values of $\rho_{th}$. From this figure we conclude that the best configuration is at $5\%$ with Option B.

%The bottom right figure shows the results for signals with higher eccentricities. We can see that the efficiency starts to decrease for signals with eccentricities bigger than 0.3.

%Sky positions sens comparison between ecliptic and poles. Also run on a frequency with a stationary line and with a moving line.
%Noise-only realizations!!Circular polarization

%\begin{figure*}[tbp]
%\centering
%\includegraphics[width=1.0\columnwidth]{ULComparisonO1.png}
%\includegraphics[width=1.0\columnwidth]{ULComparisonO1.png}
%\includegraphics[width=1.0\columnwidth]{ULComparisonO1.png}
%\includegraphics[width=1.0\columnwidth]{ULComparisonO1.png}
%\caption{\label{fig:sens}Figures showing the sensitivity of the new method. Top left: comparison with using Option B and $5\%$ of the candidates in the second step. Top right: sensitivity at three different parts of binary parameter space and frequency bands. Bottom left: comparison with different mismatch configurations. Bottom right: injections with eccentricity.}
%\end{figure*}

These tests have been done with O1 Advanced LIGO data, using data from two different detectors. The sensitivity of semi-coherent methods improves with longer observation times and by using data from more detectors, so these results should be placed in this context. Furthermore, we have analyzed four sensitivity depth points, due to the high computational cost of doing more simulations, which provide only a partial estimation of the full dependence of efficiency with sensitivity depth. Extrapolating from the results presented, we can argue that the difference in efficiency between different run configurations grows wider as the sensitivity depth is increased (i.e. as the amplitude of the signal gets smaller), but the pipeline seems to have a minimum $95\%$ efficiency at $\mathcal{D}=14$ Hz$^{-1/2}$ for all the different tests that we have done. %i.e. for the same noise floor, a run with the same duty cycle and double observing time would be X times more sensitive.

\begin{figure*}[tbp]
\centering
\includegraphics[width=1.0\columnwidth]{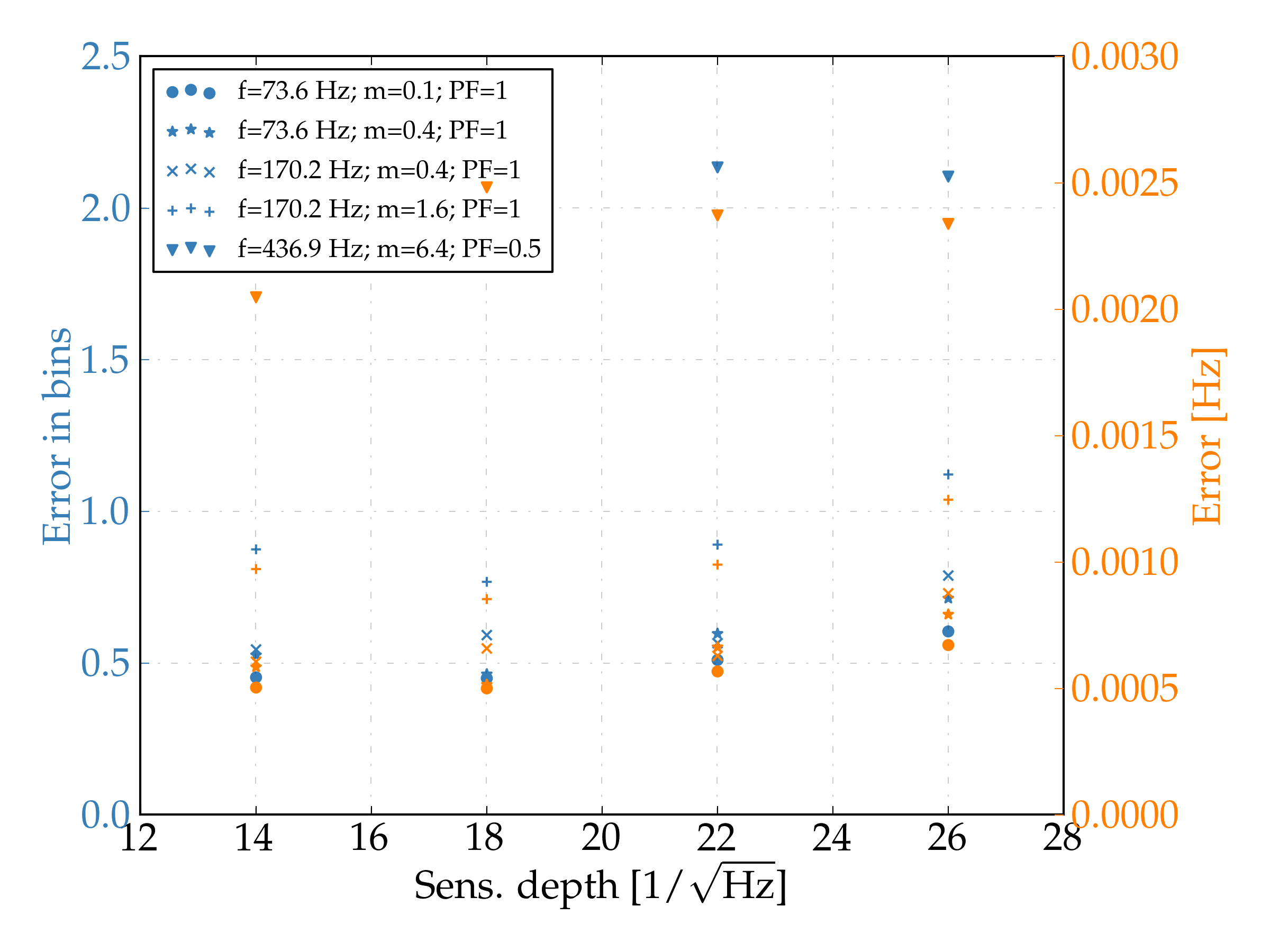}
\includegraphics[width=1.0\columnwidth]{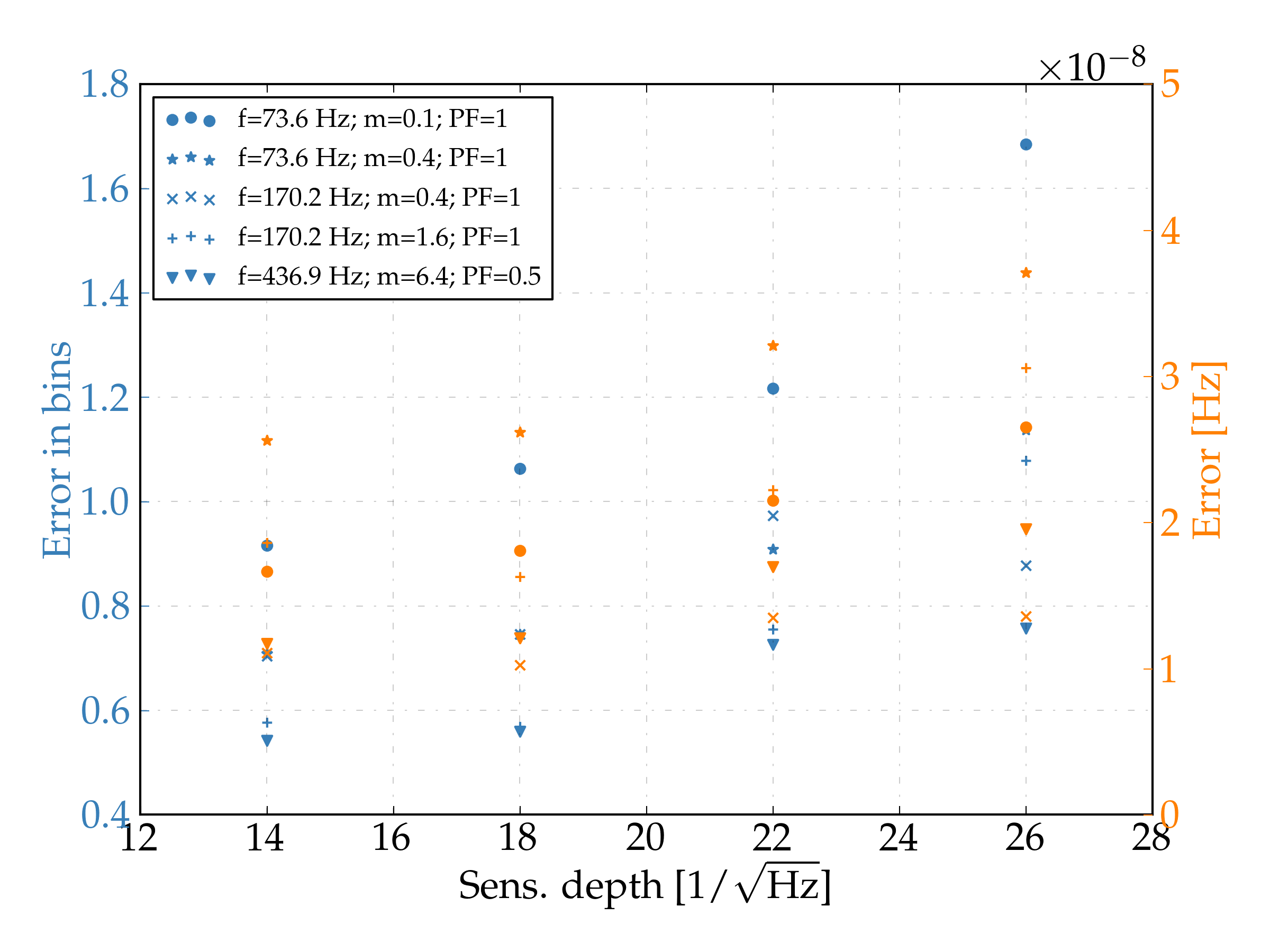}
\includegraphics[width=1.0\columnwidth]{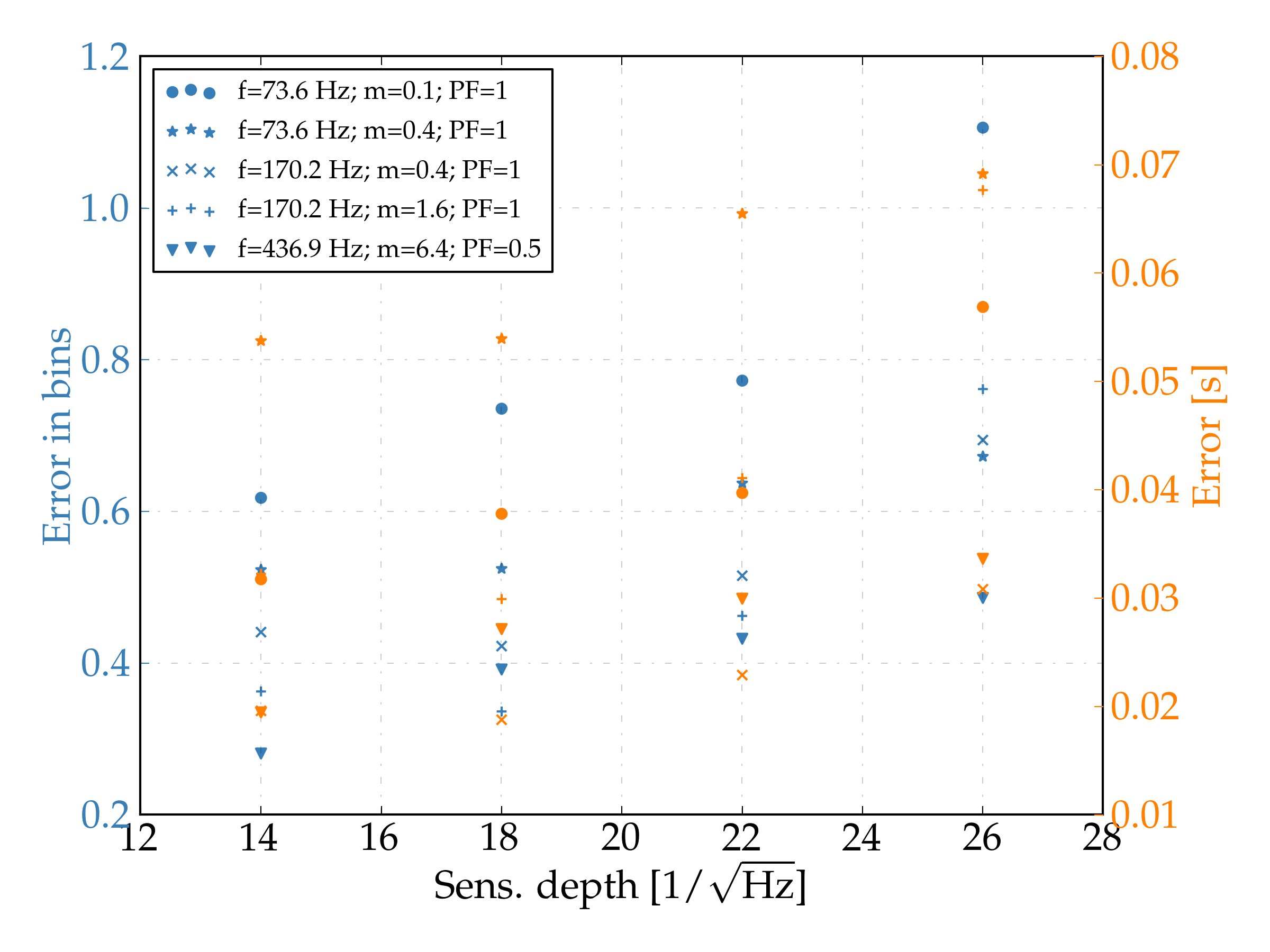}
\includegraphics[width=1.0\columnwidth]{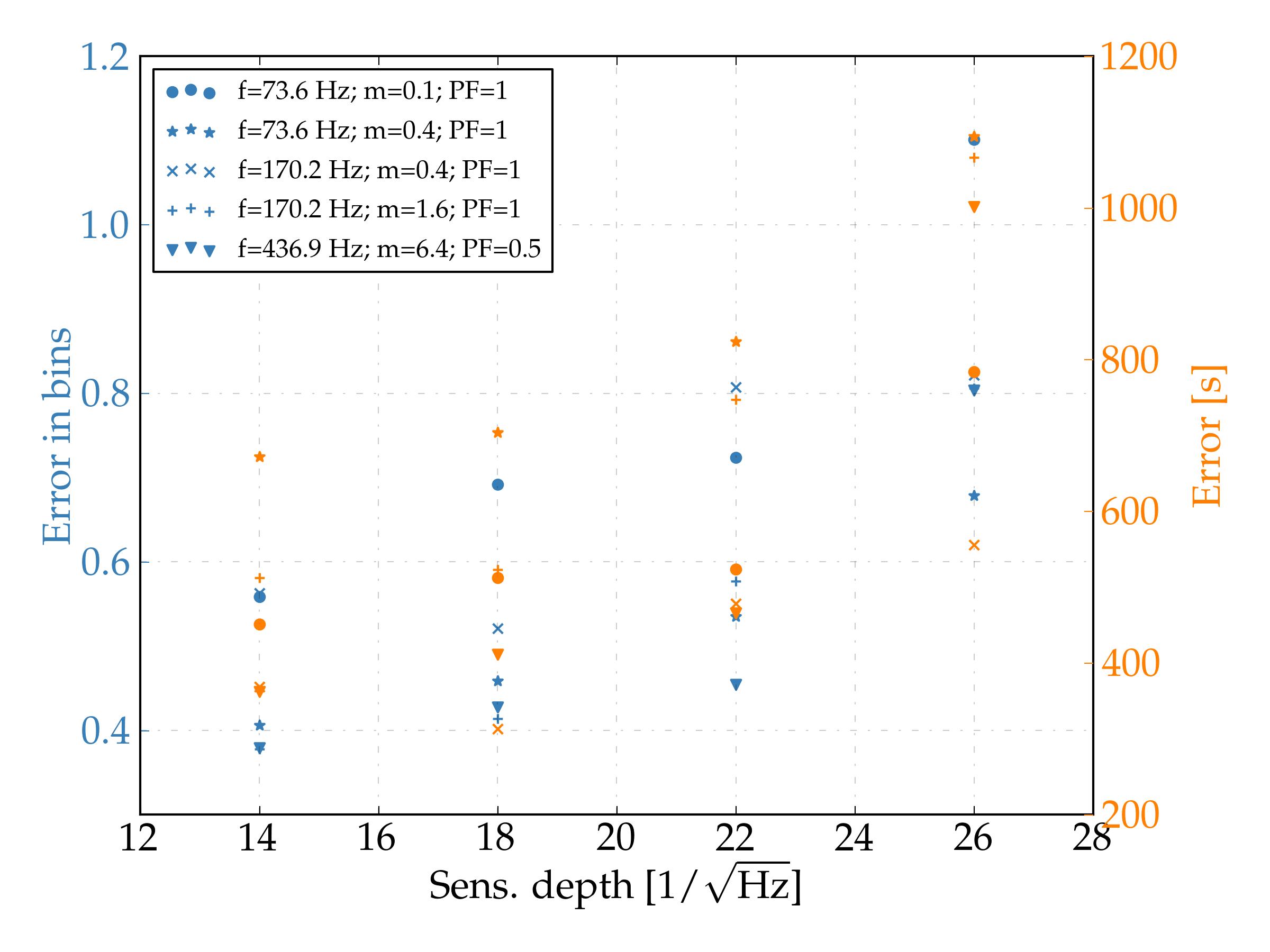}
\caption{\label{fig:ParEst}Figures showing the parameter estimation of the new method for simulations in the binary space 2, with option B and $N_C=0.05$. Top left: frequency. Top right: angular frequency $\Omega$. Bottom left: $a_p$. Bottom right: $t_{\text{asc}}$. The left vertical axis shows the mean absolute error of detected signals in number of bins, while the right vertical axis shows the mean absolute error of detected signals in the correspondent units.}
\end{figure*}

Figure \ref{fig:ParEst} shows some examples of the parameter estimation that this pipeline can achieve by comparing different mismatch configurations. It shows the results for detected simulations from the binary space 2. The errors in parameter estimation are estimated as the mean of the absolute value difference between the final cluster parameters and their true value for each injection. We observe that the different parameters show different behaviour: the configuration run with the worst estimation is not the same for all parameters. For the binary parameters, it is interesting to notice that the worst mismatch configuration usually has the best parameter estimation, both in bins and in natural units. This might be related to the frequency: for higher frequencies, the binary modulation becomes wider and the parameters can be better estimated. Comparing runs at the same frequency but with different mismatch (between orange circles and stars, or orange crosses and sums), it can be seen that the run with highest mismatch is always above the run with lower mismatch, as it should be.

\subsection{\label{sec:comparison}Comparison with other methods}

With the previous results we can translate from the sensitivity depth points at which the efficiency is $95\%$ to the estimated $h_0^{95\%}$ sensitivity. As discussed previously, for all different configurations the $95\%$ sensitivity depth is always at least at 14 Hz$^{-1/2}$, so we will take this as our sensitivity. %We can take the two extreme cases of the bottom left figure to show a range of sensitivity. This result is shown in figure \ref{fig:senzitivitycompO1}. This figure also shows a comparison with the obtained O1 \textit{SkyHough} sensitivity \cite{O1CWAllSkyLowFreq}. The difference between both sensitivities is produced by three different effects:

We can compare this result with the \textit{SkyHough} result for the O1 analysis, which for the low-frequency range (from 50 to 475 Hz) was around $24$ Hz$^{-1/2}$ \cite{O1CWAllSkyLowFreq}. The difference between these two sensitivities can be explained by these facts:
\begin{itemize}
    \item The \textit{SkyHough} search used a coherent time of 1800 s, which is twice as long as what we have used in this analysis. Longer coherent times can be used for searches for isolated systems since the frequency derivative is usually smaller. Doubling the coherent time can produce an increase in sensitivity up to approximately $1.2$, which explains a fraction of the difference. 
    \item The search for NS in binary systems has many orders of magnitude more templates. This increases the maximum values of the detection statistic for the background distribution (only noise), which has the effect of raising the thresholds in detection statistic needed to have the same false alarm rate. This effect produces a decrease of the search sensitivity, since a signal which generates the same detection statistic may be detected in one search but not in the other.
    \item The isolated search was run with $P_F=2$, while we have used $P_F=1$. This could also explain a fraction of the difference in sensitivities.
\end{itemize}

The \textit{TwoSpect} method described in \cite{TwoSpectMethods} is the only pipeline which has been used in an all-sky search for neutron stars in binary systems. From published results of a search with S6 data \cite{TwoSpectResults}, we can estimate a sensitivity depth at $95\%$ efficiency of 5 for isotropically oriented neutron stars. Taking into account the improvements developed in \cite{CohSFT} and the difference in observation time between S6 and O1, an estimated sensitivity depth with O1 data is around 6.5 Hz$^{-1/2}$, which is approximately half as sensitive as our pipeline. Furthermore, with a high number of binary blocks (e.g. around $500$) our pipeline is able to cover a similar parameter space as the one that the \textit{TwoSpect} pipeline has covered in the mentioned search, with a comparable computational cost. %The expected sensitivity for Advanced LIGO O1 of this method is $\sim7$ Hz$^{-1/2}$ \cite{TwoSpectO1}, which is approximately half as sensitive as our pipeline. Furthermore, with a high number of binary blocks (e.g. $\sim500$) our pipeline is able to cover a similar parameter space as the one that the \textit{TwoSpect} pipeline has covered in previous analyses, with a comparable computational cost. %A thorough comparison also taking into account the computational cost, the parameter space which can be analyzed and the parameter estimation is out of the scope of this paper. %can be seen in figure X \cite{TwoSpectO1}, and its computational cost is Y. A disadvantage of this method is that it only returns two binary parameters (it does not return information about the time of ascension). For this reason, a sensitive follow-up method like the one described previously is not as easily applicable as it is in our case. This method employs a non-gridded approach at its first step, and this is the reason why it loses so much sensitivity. "By reducing the binary orbital search parameters from five for the general binary orbit (three for a circularized orbit) to the two parameters used by \textit{TwoSpect}, computational efficiency is gained at the cost of sensitivity."

%The PowerFlux method firstly described and used in \cite{S4IncoherentPaper} consists on ... This is the same detection statistic that our pipeline calculates at its second stage, only four the best templates. This pipeline calculates the same but for all initial templates. We have adapted this method to work with GPUs, and we have tested its sensitivity with the same injections done before. We can see the results in figure X, where we compare both methods with the same mismatch parameters and the same data. We can see that the PowerFlux method is $X\%$ more sensitive as it was expected. The computational cost of this algorithm is X times higher than the \textit{BinarySkyHough}. We also have done tests with a GPU implementation of a power-based search (without the threshold and substitution to 1s and 0s) similar in spirit to (cite StackSlide or powerflux). This method is more sensitive, but it has a greater computational cost. We will compare the sensitivity in section ..., but the computational cost comparison is: We can see that this method takes ~5 more times to run the same parameter space.

%\begin{figure}[tbp]
%\includegraphics[width=1.0\columnwidth]{ULComparisonO1.png}
%\caption{Expected $h_0$ sensitivity at $95\%$ confidence to random polarisated signals with Advanced LIGO O1 data compared to the \textit{SkyHough} all-sky search for isolated NS for O1 data.}
%\label{fig:senzitivitycompO1}
%\end{figure}

Recently, an adaptation of the radiometer search for all-sky searches has been proposed \cite{AllSkyRadiometer}. This unmodeled search looks for coherences between two or more detectors, by tracking a signal which is inside a single frequency bin all the time. This search has a very cheap computational cost, but because it is unmodeled and the frequency bins are much coarser ($1$ Hz, as compared to our $1/900$ Hz bins), their sensitivity is worse than our pipeline. In \cite{AllSkyRadiometer}, a sensitivity of $1.2 \times 10^{-24}$ at a frequency of 245 Hz for an O1 search is quoted (with a $90\%$ confidence, compared to our $95\%$, for signals with circular polarisation and by using Gaussian data instead of more realistic data (i.e. from an observing run). These facts make a direct comparison difficult, but we can convert this value to a sensitivity depth, and try to make a rough comparison. By dividing the quoted $1.2 \times 10^{-24}$ value by the amplitude spectral density at 245 Hz, we get a value of $7$ Hz$^{-1/2}$. A realistic factor to convert this estimation to one with $95\%$ confidence, from isotropically oriented neutron stars and with realistic data is complicated to calculate, but it can be seen that our pipeline still remains at least twice as sensitive.

\begin{figure}[tbp]
\includegraphics[width=1.0\columnwidth]{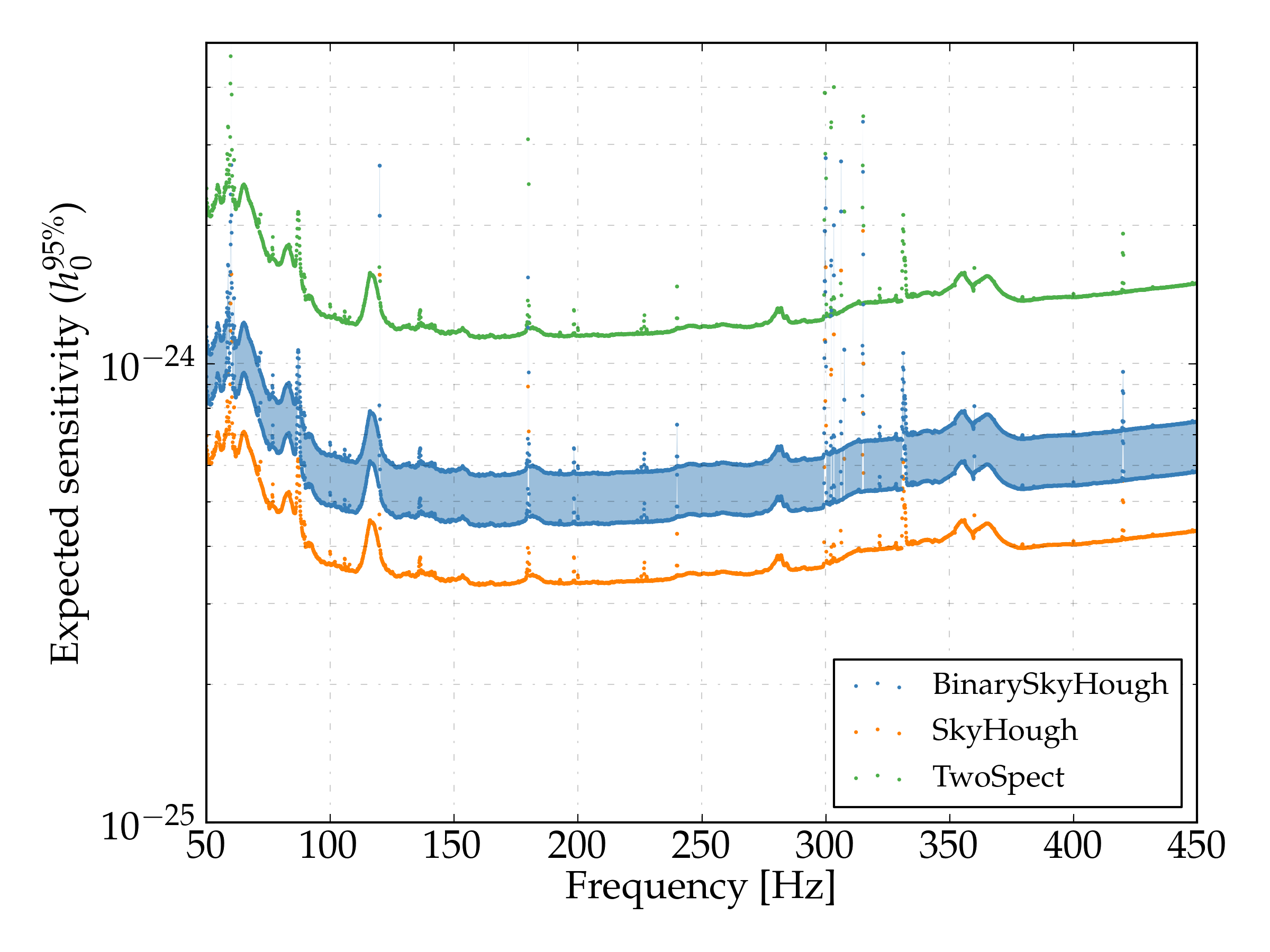}
\caption{Estimated $h_0$ sensitivity at $95\%$ confidence to random polarisated signals with Advanced LIGO O1 data compared to the \textit{SkyHough} all-sky search for isolated NS for O1 data and to the \textit{TwoSpect} pipeline.}
\label{fig:senzitivitycompall}
\end{figure}

These comparisons are shown in figure \ref{fig:senzitivitycompall} (only \textit{TwoSpect} is shown). We remark that improving the sensitivity by two means that we are able to detect signals from systems twice as far away as before, or from neutron stars with asymmetries two times smaller at the same distance. A comparison of the parameter estimation between the different pipelines has not been possible and we leave it for future work. % Comparison of parameter estimation. % covering a volume roughly 8 times larger than before (only assuming isotropic galaxy)

It is also interesting to notice the difference between \textit{BinarySkyHough} and other methods designed to perform a directed search (known sky position) for a signal from a NS in a binary system. From the published O1 results \cite{O1DirectedBinary}, we estimate a sensitivity depth of 30  Hz$^{-1/2}$, which is approximately twice our sensitivity. Since these methods don't have to search for different sky positions, the computational power can be spent in using methods which can increment the coherent time or in decreasing the mismatch parameters for the same coherent time.

\section{\label{sec:conclusions}Conclusions}
We have described a new method to perform all-sky searches of continuous gravitational waves from neutron stars in binary systems. This method is an extension of the \textit{SkyHough} pipeline, which has been coded to take advantage of GPU cards in order to overcome the prohibitive computational costs that this pipeline would have by only using CPUs, as demonstrated in section \ref{sec:compcost}.

Simulations indicate that this new method is at least $2$ times more sensitive than previous pipelines, at a comparable computational cost. It can be used to search for signals in any part of the binary parameter space, showing comparable sensitivity at all locations of parameter space. With the simulations done, it seems that regardless of the mismatch parameter chosen, the pipeline has a minimum sensitivity depth of 14 Hz$^{-1/2}$ (for a coherent time of 900 s and an observation time equal to O1). %Our method is the first method which will use a templated search from the beginning, being this the reason why it is more sensitive. This is only possible with the usage of massive parallelisation (e.g. by using GPUs), which allows the completion of the run in a feasible time. Furthermore, our method returns the three binary parameters that are needed to do follow-up studies of possible candidates with more sensitive methods like F-statistic, a feature which TwoSpect does not share.

%There is plenty of room for improvement of this pipeline.  Furthermore, cleaned data could be used and better background estimation, using data combined from two detectors, using tau statistic, ... 

A possibility to improve the sensitivity of this pipeline would be to use cleaned data as input data. Some procedures can remove time-domain disturbances which affect some of the generated SFTs, and these cleaned SFTs can improve the sensitivity of a search with no additional computational cost, as shown in \cite{FreqHough}. Another improvement in sensitivity could come from the generation of modified SFT bins, which take into account the leakage of power due to the signal not being at exactly the center of a frequency bin \cite{AllenPapa}. Furthermore, coherently combining data from different detectors as explained in \cite{CohSFT} may also improve the sensitivity of the pipeline.

%The computational cost of the pipeline could be diminished by running simultaneously in one GPU card two or more CPU jobs (analyzing different bands) at the same time. The code switches between CPU and GPU calculations, and some times the GPU is sitting idle. By running more than one job at the same time this could be avoided, and the time needed to complete a search could be reduced. %We leave this characterizations for future work. % Also better lattices.

An optimal way of spending a limited computation budget should be developed in order to maximize the chances of detecting a signal. An analytical estimation of how each parameter of the pipeline affects the sensitivity and the computational cost would need to be estimated, but this development would increase the possibilities of detection.%, e.g. is it better to increase the coherent time, to decrease the mismatch or to increase the number of templates which are recalculated in the second step?

%A more in-depth study of the binary metric and mismatch for all-sky semi-coherent searches is left for future work. Also windows for post-processing.

%We have shown a possible follow-up method using an MCMC search, which can be used to increase the significance of any outlier and enhance the parameter estimation.

The \textit{BinarySkyHough} pipeline could also be used to perform the first all-sky search which explicitly looks for NS in high-eccentricity systems. Two more parameters ($e$ and $\omega$) should be included, and this would further increase the computational cost of the search, narrowing the range of parameter space which could be searched. This pipeline (with some modifications) could also be applied for a directed search of a NS in a binary system, where the sky position is known but the frequency and binary parameters are unknown. By eliminating two parameters of the search, the computational cost could be used for analyzing a broader frequency or binary range or to use lower mismatch parameters.

We plan to keep developing this pipeline to prepare it to analyse the upcoming O3 observing run, and we also plan to apply it to a search using data from the O2 Advanced LIGO observing run.

%Conclusions:
%    Most sensitive search
%    Possibility of sensitive follow-up

%Outlook:
%    Test search, O3
%    NS in eccentric binary systems

%Apply this method or PowerFLux method to directed searches like Sco X-1
    
%Running isolated all-sky along with binary all-sky: no extra cost Downside: less Tcoh -> only possible for highest frequencies

\begin{acknowledgments}
The authors want to thank Sharon Brunnett for helping with the computational setup and compilation of the new code, and Evan Goetz and John Whelan for comments and suggestions on a draft of this paper. We acknowledge the support of the Spanish Agencia Estatal de Investigaci{\'o}n and Ministerio de Ciencia, Innovaci{\'o}n y Universidades grants FPA2016-76821-P, FPA2017-90687-REDC, FPA2017-90566-REDC, FPA2015-69815-REDT, FPA2015-68783-REDT, the Vicepresidencia i Conselleria d'Innovaci{\'o}, Recerca i Turisme del Govern de les Illes Balears and the Fons Social Europeu 2014-2020 de les Illes Balears, the European Union FEDER funds, and the EU COST actions CA16104, CA16214 and CA17137. The authors are grateful for computational resources provided by the LIGO Laboratory and supported by National Science Foundation Grants PHY-0757058 and PHY-0823459. This article has LIGO document number P1900091.
%Greg Ashton for follow-up. Josh Willis for optimizing the code.
\end{acknowledgments}

\end{document}